\documentclass[%
 twocolumn,
 reprint,
 float,
 pst-func,
showkeys,
superscriptaddress,
 amsmath,amssymb,
 aps,
 pre,
bbm,
float
]{revtex4-1}

\usepackage{xcolor}

\usepackage{comment}

\usepackage{graphicx}
\usepackage{dcolumn}
\usepackage{bm}
\usepackage{hyperref}

\begin{document}

\preprint{}

\title{Statistical properties of avalanches via the $c$-record process}

\author{Vincenzo Maria Schimmenti}
\author{Satya N. Majumdar}%
\author{Alberto Rosso}
\affiliation{
Universit\'e  Paris-Saclay,  CNRS,  LPTMS,  91405,  Orsay, France
}%

\date{\today}

\begin{abstract}

We study the statistics of avalanches, as a response to
an applied force, undergone by a particle
hopping on a one dimensional lattice where the
pinning forces at each site are
independent and identically distributed (I.I.D),
each drawn from a continuous $f(x)$. The avalanches in
this model correspond to the inter-record intervals
in a modified record process of I.I.D variables, defined
by a single parameter $c>0$. This parameter characterizes
the record formation via the recursive process
$R_k > R_{k-1}-c$, where $R_k$ denotes the value of the $k$-th record.
We show that for $c>0$, if $f(x)$ decays slower than an exponential for 
large $x$,
the record process is nonstationary as in the standard $c=0$ case.
In contrast, if $f(x)$ has a faster than exponential tail, the
record process becomes stationary and the avalanche size distribution
$\pi(n)$ has a decay faster than $1/n^2$ for large $n$. The
marginal case where $f(x)$ decays exponentially for large $x$
exhibits a phase transition from a non-stationary phase
to a stationary phase as $c$ increases through a critical
value $c_{\rm crit}$. Focusing on
$f(x)=e^{-x}$ (with $x\ge 0$), we show that 
$c_{\rm crit}=1$ and for $c<1$, the record
statistics is non-stationary. However, for $c>1$, the
record statistics is stationary with avalanche size distribution
$\pi(n)\sim n^{-1-\lambda(c)}$ for large $n$. Consequently, for $c>1$,
the mean number of records up to $N$ steps grows
algebraically $\sim N^{\lambda(c)}$ for large $N$. 
Remarkably, the exponent
$\lambda(c)$ depends continously on $c$ for $c>1$ and is given
by the unique positive root of $c=-\ln (1-\lambda)/\lambda$. We also unveil
the presence of nontrivial correlations between avalanches in the
stationary phase
that resemble earthquake sequences.

\end{abstract}

\maketitle

\section{\label{sec:intro}Introduction}

Records are ubiquitous in nature. We keep hearing about record breaking events in 
sports, in stock prices, in the summer temperature in a given city, in 
the amount of rainfall in a given place or in the magnitude of 
earthquakes in a certain geographical zone. The studies on the theory of 
records were initiated in the statistics literature almost 70 years 
back~\cite{Chandler52,FS54,Neuts67,Scho72,ABN98,Nevzorov} and since then have found 
numerous applications across disciplines: in 
sports~\cite{Gembris2002,Gembris2007,BRV2007}, in the analysis of climate 
data~\cite{Hoyt81,Basset92,SZ1999,RP06,Meehl2009,WK,AK,MBK19}, in 
fitness models of evolutionary 
biology~\cite{KJ,KRUG07,PSNK15,PNK16,PK16}, in condensed matter systems 
such as spin glasses and high temperature 
superconductors~\cite{Jensen06,Oliveira2005,Sibani2006} and also in 
models of growing networks~\cite{GL1}. Record statistics have also been 
studied extensively in various random walk 
models~\cite{MAJ08,SMreview,SS,MSW12,EKMB13,GMS15,GMS16,MOUNAIX20} with 
applications to avalanches and depinning of elastic lines in disordered 
medium~\cite{LDW09}, to the analysis of financial 
data~\cite{WBK11,WMS12,SS14,Chalet17} and more recently to active 
particles with run and tumble dynamics~\cite{MLMS20,LM20,MLMS21}.
For reviews on record statistics in the physics literature, see
Refs.~\cite{Wergen13,GMS17}.

\begin{figure}
\centering
\includegraphics[width=\linewidth]{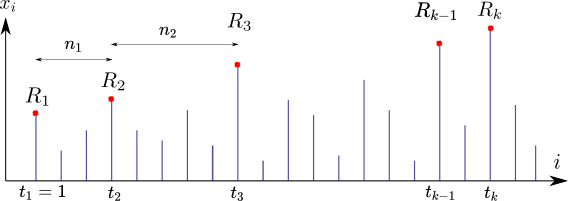}
\caption{An infinite discrete-time series with entries 
$\{x_1,x_2,\cdots\}$. A record happens at step $k$ if
$x_k>x_i$ for all $i=1,2,\cdots, k-1$. 
The successive record values (shown by (red) filled circles)
$\{R_1,\,R_2,\, R_3,\,\cdots\}$ form the \emph{record-series}. The times at
which the records occur form a \emph{record-time} series $\{t_1,\,t_2,\,t_3,\,
\cdots\}$.
The time gap $n_k= t_{k+1}-t_k$ between the $k$-th record
and the $(k+1)$-th record is called the \emph{age} of the $k$-th
record. By convention, the first entry is a record, hence $R_1=x_1$
and $t_1=1$.} 
\label{fig:sequence1}
\end{figure}

In its most general setting, the record problem can be formulated as
follows. Consider an infinite sequence of continuous random variables
$\{x_1,\,x_2,\,x_3,\ldots\}$ representing the entries of
a discrete-time series--they may be the stock prices on successive days
or the daily average temperature in a given city. The random variables
are distributed via a joint distribution $P(x_1,\,x_2,\,x_3,\ldots)$.
A record (upper) occurs at step $k$ if the entry $x_k$ exceeds all
previous entries, i.e., if
\begin{equation}
x_k> x_i, \quad {\rm for}\,\, {\rm all}\quad i=1,2,\cdots, k-1 \, .
\label{def_record}
\end{equation}
By convention, the first entry is always a record. 
The successive record
values are denoted by $\{R_1,\, R_2,\, R_3,\, \cdots\}$ and
is called the associated {\em record-series} (see Fig. (\ref{fig:sequence1})).
Furthermore, let $\{t_1,\,t_2,\, t_3,\cdots\}$ denote the times at
which the records occur--we will call it the associated 
{\em record-time series}. Since the first entry is always a record
by convention, we have $R_1=x_1$ and $t_1=1$. 

Given this time-series
$\{x_1,\, x_2,\, x_3,\cdots\}$ and its underlying
probability distribution $P(x_1,\, x_2,\, x_3,\cdots)$, one can
investigate various observables associated to the occurrences
of records. Three rather natural questions are as follows:

\begin{itemize}

\item How many records $M_N$ occur in the first $N$ steps? For example,
given the joint distribution $P(x_1,\, x_2,\, x_3,\cdots)$, what is
the average number of records $\langle M_N\rangle$ within the first $N$
steps?

\item How are the record values distributed? For example, let
\begin{equation}
q_k(R)= {\rm Prob.}[R_k=R]
\label{def_qkr}
\end{equation}
denote the probability density that the $k$-th record (in the infinite sequence)
takes value in $[R, R+dR]$. What can we say about $q_k(R)$, given
the underlying joint distribution $P(x_1,\, x_2,\, x_3,\cdots)$ of the
original entries?

\item Suppose that a record occurs at time $t_k$. How long does it
take to break this record? Let $n_k= t_{k+1}-t_{k}$ denote the
time gap between the $(k+1)$-th record and the $k$-th record--this
is the {\em age} of the $k$-th record. Let
\begin{equation}
\pi_k(n)= {\rm Prob.}[n_k=n]
\label{def_pikn}
\end{equation}
denote the distribution of the age of the $k$-th record. Given
the joint distribution of entries $P(x_1,\, x_2,\, x_3,\cdots)$,
can one compute $\pi_k(n)$? 

\end{itemize}

\vskip 0.4cm

\noindent {\bf I.I.D model.} The simplest model where one can compute exactly all three observables
correspond to the case when the entries $x_i$'s are {\em uncorrelated}
and each is drawn independently from a continuous distribution $f(x)$.
In other words, the joint distribution factorizes
\begin{equation}
P(x_1,\, x_2\, x_3,\cdots)= f(x_1)\, f(x_2)\, f(x_3)\cdots
\label{factor.1}
\end{equation}
This is usually referred to as the independent and identically distributed
(I.I.D) model~\cite{Chandler52,FS54,Neuts67,Scho72,ABN98,Nevzorov}. 
The probability density function (PDF) $f(x)$ is
normalized to unity and its cumulative distribution is defined as
$F(x)=\int_{-\infty}^x f(y)\, dy$. We summarize these classical 
results here, and for a derivation see, e.g., the review~\cite{GMS17} with
citations to the original 
literature~\cite{Chandler52,FS54,Neuts67,Scho72,ABN98,Nevzorov}. 

\vskip 0.3cm

\noindent (i) It turns out that the average
number of records up to first $N$ steps is universal, i.e., independent
of $f(x)$ and is given by the simple formula
\begin{equation}
\langle M_N\rangle= \sum_{k=1}^N \frac{1}{k} \xrightarrow[N\to \infty]{} 
\ln N\, . 
\label{avg_rec.1}
\end{equation}
Thus the mean number of records grows very slowly (logarithmically) with
increasing $N$ indicating that records get increasingly harder to break.
Moreover, the full distribution of $M_N$, i.e., $P(M,N)={\rm Prob.}(M_N=M)$
is also known and is universal. In the limit $N\to \infty$, the
distribution $P(M,N)$ approaches a Gaussian form
with mean $\ln N$ and variance $\ln N$. 

\vskip 0.3cm

\noindent (ii) The distribution $q_k(R)$
of the value of the $k$-th record is also known explicitly
\begin{equation}
q_k(R)= f(R)\, \frac{\left[-\ln (1- F(R))\right]^{k-1}}{(k-1)!}\, ,  
\quad k=1,2,\ldots\, .
\label{qkr_iid.1}
\end{equation}
where $F(R)$ is the cumulative distribution. Unlike the distribution
of $M_N$, the record value distribution $q_k(R)$ is not universal, as
it depends explicitly on $f(R)$. For example, for 
exponentially distributed positive entries with $f(x)= e^{-x}\,\theta(x)$
(where the Heaviside step function $\theta(x)=1$ if $x>0$ and 
$\theta(x)=0$ if $x\le 0$),
Eq. (\ref{qkr_iid.1}) gives
\begin{equation}
q_k(R)= e^{-R}\, \frac{R^{k-1}}{(k-1)!}\,   \quad k=1,2\cdots
\label{qkr_exp_iid.1}
\end{equation}
In this case the average record value $\langle R_k\rangle= \int_0^{\infty}
R\, q_k(R)\, dR= k$ increases linearly with $k$. Similarly, the variance
of $R_k$ also grows linearly with $k$. In fact, for generic $f(x)$
in Eq. (\ref{qkr_iid.1}), one can show that $q_k(R)$ does not
have a limiting distribution as $k\to \infty$. 

\vskip 0.3cm

\noindent (iii) Finally, the age distribution
$\pi_k(n)$ of the $k$-th record is also known explicitly and
turns out to be universal, i.e., independent of $f(x)$. For
any $k\ge 1$, it reads~\cite{GMS17}
\begin{equation}
\label{pikn_iid.1}
     \pi_k(n)=\sum_{m=0}^{n-1}\binom{n-1}{m} \frac{(-1)^m}{(2+m)^k}  
\underset{n \to \infty}{\simeq} \frac{1}{n^2}\,
\frac{\left[\ln n\right]^{(k-1)}}{(k-1)!}\, .
\end{equation}
Two important points to note for the I.I.D model that will
be important in this paper: (a) the distribution $q_k(R)$
depends on $k$ even when $k\to \infty$, i.e.,
there is no limiting {\em stationary} record value distribution as
$k\to \infty$, simply because the record
values grow with $k$ for generic $f(x)$ and (b) 
similarly, the age distribution $\pi_k(n)$ in Eq. (\ref{pikn_iid.1})
also does not have a limiting stationary distribution when $k\to \infty$.

\vskip 0.3cm

After introducing this basic background, we now turn to the main topic 
of this paper. Here our goal is to use the tools of record statistics to 
provide a simple model for the peculiar jerky motion observed in many 
disordered systems, as a response to an external driving force. 
In such systems the dynamics of the relevant degree of freedom 
typically alternates between static immobile 
states and periods of rapid motion called avalanches. Examples of such a 
behaviour are quite ubiquitous, ranging from the crack propagation in 
solids \cite{BB11,Bonamy17,BBR19}, the earthquake of seismic faults 
\cite{AGGL16} or the Barkhausen noise \cite{ABBM90,ZVS97,SDM01} appearing 
in the magnetisation curve as a function of the applied magnetic 
field--see \cite{FC2008} for a review. Avalanches are well studied in 
the context of the depinning of an elastic interface pulled through
a disordered medium by an external force~\cite{LDW09,ABBM90,Kardar98,Fisher98}. 
In the absence 
of an external force the line is pinned by the disorder. Upon increasing 
the force beyond a local threshold, a portion of the interface gets 
depinned and its center of mass moves forward, thus creating an 
avalanche. Interestingly, a simple one dimensional lattice 
model for depinning can 
be mapped exactly to the record model discussed above, as we show below.

In this lattice model,
the elastic line is replaced by a single particle (representing
its center of mass) that moves on an infinite $1$-d lattice.
Under this mapping, the time series $\{x_i\}$ in Fig. (\ref{fig:sequence1})
gets mapped on to the quenched pinning forces with the horizontal
axis $i$ labelling the sites of a $1$-d lattice. 
This defines the disorder landscape $\{x_1,x_2,\dots \}$ on which the 
particle moves under the effect of the applied force, 
$f_a(i)$ (namely the force applied at the site $i$). 
The particle leaves the site $i$ if $f_a(i) > x_i$. 
Hence, the minimal force profile allowing  motion 
alternate between plateaus and vertical jumps (see Fig.2 ). 
The plateaus coincide exactly with the record series 
$\{R_1,\,R_2,\,R_3,\dots\}$ and the vertical jumps occur 
exactly at the sites 
$\{t_1,\,t_2,\, t_3 \dots\}$ where the record occurs in Fig.1. 
The age of the $k$-th record $n_k$
in Fig. (\ref{fig:sequence1}) maps on to the size of the $k$-th avalanche
in the depinning model. The three observables $\langle M_N\rangle$,
$q_k(R)$ and $\pi_k(n)$ have precise physical meaning in
the context of depinning. For example, $\langle M_N\rangle$
is the number of jumps of the applied force profile in
order to displace the position of the particle by $N$ sites, given
that it started at $i=1$. Similarly, $\pi_k(n)$ represents
the size distribution of the $k$-th avalanche in the depinning
model. 

In the simple I.I.D setting, the pinning forces $\{x_i\}$
in the disordered landscape are independent and identically
distributed, each drawn from a continuous PDF $f(x)$. However,
as we discuss in detail in Section (\ref{sec:depinning}), 
this simple I.I.D model, while analytically tractable, 
fails to reproduce the behaviors of the
three observables as seen in real systems. In addition,
the spatio-temporal correlations in the
applied force profile $f_a(i)$ (record values) as well as in the avalanches
seen in realistic systems are also not reproduced by the I.I.D model.
This calls for some amendments in this basic I.I.D model. 
The idea is to introduce minimal changes
in the model such that it retains its analytical tractability, and
yet reproduces the features observed in real systems. 

After discussing briefly the previous attempts of modifying the simple
I.I.D model in Section \ref{sec:depinning}, 
we introduce a new model in this paper which we
call the $c$-record model. This model has the I.I.D landscape
with input $f(x)$, and one single additional parameter $c>0$
associated with the applied force profile $f_a(i)$. 
The model is precisely defined in Section \ref{sec:record_process}.
We demonstrate in this paper that the the $c$-record model is 
(a) exactly solvable
with a rich analytical structure and (b) it reproduces
qualitatively similar features for all the three observables,
as well as the spatio-temporal correlations, that one observes
in realistic system of depinning.

Quite remarkably, it turns out that this $c$-record model was
already introduced in a different context in the statistics literature
by Balakrishnan et. al., where is was called the $\delta$-{\it exceedence}
record model~\cite{Bala96}. The parameter $\delta=-c<0$ is negative
in our context. In addition, this $\delta$-{\em exceedence} model
with a negative $\delta=-c<0$ also appeared in the {\em random 
adaptive walk} (RAW) model to describe biological evolution
on a random fitness landscape~\cite{PSNK15,PNK16,PK16}.
In fact, we use the notation $c$ for $-\delta$ in our model
following Ref.~\cite{PSNK15}. 

In the context of the RAW model,
the two observables $\langle M_N\rangle$ and $q_k(R)$, but not
$\pi_k(n)$, were
already studied analytically in Ref.~\cite{PSNK15}. 
In the notation of Ref.~\cite{PSNK15} for RAW, our
$\langle M_N\rangle$ corresponds to the mean walk length
of an adaptive walker (for genome size $L$) $D_{\rm RAW}(L)$ 
in the RAW model, with $L\sim N$ for large $L$.
In particular,
for exponentially distributed positive fitness landscapes
$f(x)= e^{-x}\theta(x)$ where $\theta(x)$ is the Heaviside step function, 
Ref.~\cite{PSNK15} uncovered a striking
phase transition at the critical value $c=1$, across which
the asymptotic growth of $\langle M_N\rangle$, with
increasing $N$, changes drastically.

In this paper, we re-visit this $c$-record model
in the context of depinning and avalanches.
We provide a thorough analytical and numerical study of
all three observables $\langle M_N\rangle$, $q_k(R)$ and
$\pi_k(n)$ and also the underlying correlation structure
of the record values and avalanches for a general $f(x)$, 
including in particular the interesting
case $f(x)=e^{-x}\theta(x)$. For this particular
case $f(x)=e^{-x}\theta(x)$, while our results fully
agree with Ref.~\cite{PSNK15} for $c\le 1$, we show
that for $c>1$ the model has a much richer structure than
was reported in Ref.~\cite{PSNK15}. In particular, we show that for $c>1$ and
$f(x)=e^{-x}\theta(x)$,
the average number of records $\langle M_N\rangle$ grows for large
$N$ as a power law
\begin{equation}
\langle M_N\rangle \sim N^{\lambda(c)}\, ,
\label{avg_rec.intro}
\end{equation}
with an exponent $\lambda(c)$ that depends continuously on $c$ (for $c>1$)
and is given by the unique positive root of the transcendental equation
\begin{equation}
c= -\frac{\ln (1-\lambda)}{\lambda}\, .
\label{lambda_intro}
\end{equation}
Thus our prediction for the asymptotic growth of $\langle M_N\rangle$
in Eq. (\ref{avg_rec.intro}) for $c>1$ differs from Ref.~\cite{PSNK15}
where $\langle M_N\rangle \sim O(N)$ was reported.
We also show that for $c>1$, the record value distribution $q_k(R)$
approaches a stationary distribution as $k\to \infty$, which is
given by a pure exponential behaviour for all $R\ge 0$
\begin{equation}
q_{k\to \infty}(R)= \lambda(c)\, e^{-\lambda(c)\, R}\, .
\label{qkr_intro}
\end{equation}
In addition, the avalanche size distribution $\pi_k(n)$ also
approaches a stationary distribution as $k\to \infty$ (for $c>1$)
with a power-law tail 
\begin{equation}
\pi_{k\to \infty}(n)\sim n^{-\lambda(c)}\, \quad {\rm as}\quad n\to \infty
\label{pikn_intro}
\end{equation}
where the exponent is the same $\lambda(c)$ as in Eq. (\ref{lambda_intro}).   

The rest of our paper is organised as follows. In Section 
\ref{sec:depinning}, we recall the mapping between the $1$-d lattice model
of depinning and the record model and also discuss previously studied models
that go beyond the simple I.I.D record model. In Section \ref{sec:record_process},
we define the $c$-record model precisely, and provide a detailed
summary of our results. For the particular case $f(x)= e^{-x}\theta(x)$,
we also provide a detailed comparison of our results to that of 
Ref.~\cite{PSNK15}. In Section \ref{recursion} we set up the exact recursion
relations and derive the non-local differential equations
for the three main observables, respectively in Subsections \ref{sec:recursive_rels_MN},
\ref{sec:recursive_rels_qR}, and
\ref{sec:recursive_rels_pi}.
In Section \ref{sec:exp_case}, we provide the full exact solution of all three
observables in the $c$-record model and demonstrate the phase transition at $c=1$.
In Section \ref{stretched} we discuss in detail the criterion for stationarity
of the record value distribution $q_k(R)$ as $k\to \infty$ for a stretched exponential
family of $f(x)$. Section \ref{other_dist} considers other fanilies of $f(x)$, including
an exact solution for the uniform distribution over $[0,1]$ and numerical results
for the Weibull class of $f(x)$.  
In Section \ref{sec:generalization} we show some possible
generalizations of the $c$-record process. Section \ref{sec:conclusion}
is dedicated to the conclusion. Finally, the 
derivation of the asymptotic results, 
rather long and tedious, are relegated to Appendices (A-G).

\section{\label{sec:depinning} Depinning, avalanches and record statistics}

To understand how record
statistics of a discrete-time series, discussed in the introduction, 
can be used to study the avalanches associated with the depinning of 
an elastic interface, we consider a very simple one 
dimensional model where one replaces the extended 
interface by a point representing its center of mass~\cite{LDW09}.
The model is defined on an infinite one dimensional lattice where the 
lattice sites are labelled by $i=1,\,2,\,3,\cdots$. At each site $i$,
we assign a positive random variable $x_i$, drawn independently from
a continuous $f(x)$, representing the local pinning force at site $i$.
The time-series $\{x_i\}$ in Fig. (\ref{fig:sequence1}) then
defines the quenched random landscape, with the horizontal
axis $i$ labelling the lattice sites. The associated
record-series $\{R_1,\,R_2,\, R_3,\cdots\}$ in Fig. (\ref{fig:sequence1})
now defines the record values of this pinning force landscape.
We then launch a single
particle on this quenched landscape at site $i=1$ and apply an external force 
$f_a$ at site $i=1$ that
tries to drag the particle from $i=1$ to the neighbouring site $i=2$.
The force $f_a$ is increased continuously with time at a constant rate. 
As long as the value of $f_a$ is less than the local pinning force $x_1$,
the particle does not move from site $1$. Upon increasing the force $f_a$, 
when it just exceeds $x_1=R_1$, the particle suddenly jumps to site $2$.
Let $f_a(1)$ denote the applied force just when the particle leaves the site $i=1$.
The value of the applied force remains the same, i.e., $f_a=f_a(1)=x_1=R_1$ when
the particle is moving. When the particle arrives at site $i=2$,
if the current force $f_a(1)$ is less than $x_2$, i.e., $x_2$ is a record,
the particle gets stuck again at site $2$ and
the applied force needs to increase to exceed the pinning force $x_2$. However, if $x_2$ is not a record,
the current force $f_a(1)$ is bigger than $x_2$ and the particle hops from $i=2$ to
$i=3$. Essentially the particle keeps moving forward to the right 
till it encounters
the next record value of the landscape. We then have to increase the
force $f_a$ to exceed the current record value and the process continues.
The number of sites the particle moves forward following a depinning
(till it gets pinned again) is precisely the size of an avalanche.

\begin{figure}
\centering
\includegraphics[width=\linewidth]{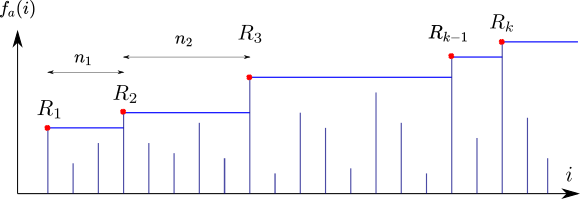}
\caption{The applied force profile $f_a(i)$ in the
$1$-d depinning model
has a staircase structure (shown by the
solid black line), alternating
between plateaus and vertical jumps. The plateau values coincide
exactly with the record-series $\{R_1,\,R_2,\,R_3\,\cdots\}$,
while the jumps occur precisely at the sites where the records
occur, i.e., at sites $\{t_1,\,t_2,\,t_3,\cdots\}$ of figure (\ref{fig:sequence1}). The
age of the $k$-the record $n_k$ coincides exactly with the size of the
$k$-th avalanche in the depinning model.} 
\label{fig:sequence2}
\end{figure}

Let $f_a(i)$ denote the value of the applied force at site $i$ 
just when the particle leaves the site $i$.
In this simple model, we thus see that the applied force profile $f_a(i)$
essentially has a staircase structure alternating between plateaus
and vertical jumps (see Fig.~(\ref{fig:sequence2})). The plateau
values of the force $f_a(i)$ are precisely the record values 
$\{R_1,\, R_2,\, R_3,\cdots\}$ of the underlying landscape and
the jumps of $f_a(i)$ occur exactly at the sites where records occur
in the quenched landscape, i.e,
they coincide with record-time series $\{t_1,\, t_2,\, t_3\, \cdots\}$
in Fig. (\ref{fig:sequence1}). Consequently, the ages $n_k$ of
the records coincide exactly with the sizes of successive avalanches.
Thus the three observables introduced before in Fig. (\ref{fig:sequence1})
are also very relevant in the context of depinning: (i) $\langle M_N\rangle$
measures the average number of jumps the external force has to undergo
in order to displace the particle from site $i=1$ to site $i=N$, (ii) $q_k(R)$
represents the distribution of the height of the $k$-th plateau
of the applied force and (iii) $\pi_k(n)$ is precisely the distribution
of the size of the $k$-th avalanche. 

How realistic is this simple depinning model with an I.I.D landscape?
It turns out that there are three important features in real systems that
the I.I.D model fails to reproduce faithfully: 

\vskip 0.3cm

\noindent (1) In real systems,
the distribution $q_k(R)$
of the height of the $k$-th plateau in the force profile typically
approaches a stationary distribution as $k\to \infty$. In contrast,
in the I.I.D model, as
seen from Eq. (\ref{qkr_iid.1}), there is no limiting
stationary distribution for $q_k(R)$ as $k\to \infty$.

\vskip 0.3cm

\noindent (2) In real systems,
the avalanche size distribution $\pi_k(n)$ not only approaches
a stationary distribution $\pi(n)$ as $k\to \infty$, but
the stationary distribution also
has a pure power-law tail, $\pi(n)\sim
n^{-\tau}$ as $n\to \infty$~\cite{ZVS97,BCDS02,BDHDB18} 
(e.g., the celebrated Gutenberg-Richter law 
for earthquake magnitude). In the I.I.D model, the result in
Eq. (\ref{pikn_iid.1}) indicates
that $\pi_k(n)$ neither approaches a stationary distribution as $k\to \infty$,
nor does it have a pure power-law tail. 

\vskip 0.3cm

\noindent (3) Real systems exhibit very interesting correlations 
between the record ages, $n_k$, as well as between the record values, 
$R_k$. For example, after a large earthquake occurs, one observes a 
cascade of large aftershocks followed by long periods of quiescent 
activity characterized by events of small size. It turns out that in the 
I.I.D model the record values $R_k$'s increase monotonically with $k$ 
and a similar trend is observed for the ages, leaving no scope for 
observing cascade of events of large sizes, followed by quiescent 
activity.

\vskip 0.3cm

An important improvement over the I.I.D model is represented
by the well known Alessandro-Beatrice-Bertotti-Montorsi (ABBM)
model introduced to study the avalanches in Barkhausen noise~\cite{ABBM90}.
In the ABBM model, the I.I.D landscape is replaced by a correlated
one where $x_i$ represents the position of a one dimensional 
random walk~\cite{ABBM90,FC2008}. In this case, the avalanche
size distribution $\pi_k(n)$ coincides with the
return time distribution of a Brownian motion in $1$-d, and
hence $\pi_k(n)$ is stationary (independent of $k$) and
does have a pure power law tail, $\pi_k(n)\sim n^{-\tau}$ with
$\tau=3/2$. A similar analysis also follows from the study
of record statistics for a random walk sequence~\cite{MAJ08}.
However, in this model, $q_k(R)$ does not approach
a stationary distribution as $k\to \infty$ (the point (1) above) 
and the sequence of record ages is uncorrelated at variance with 
what it is seen in real systems (the point (3) above).

\begin{figure}
\centering
\includegraphics[width=\linewidth]{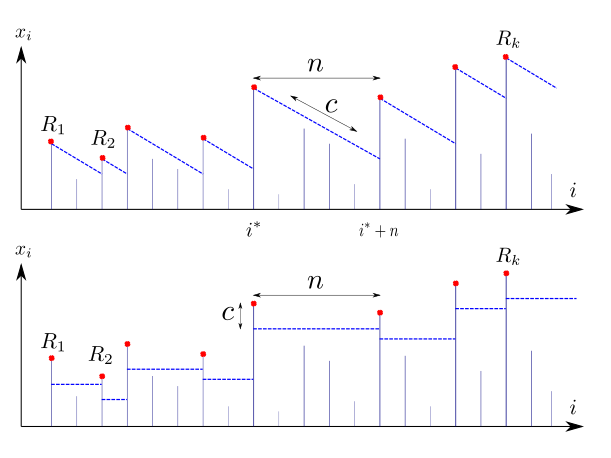}
\caption{Schemes for modified records. Upper panel: linear trend records. 
Lower panel: $c$-records.}
\label{fig:schema}
\end{figure}

Another modification of the simple I.I.D record model is the
so called linear-trend model~\cite{LDW09}. In this model,
the landscape of pinning forces $\{x_i\}$ remains I.I.D, but
the applied force profile $f_a(i)$ changes from Fig. (\ref{fig:sequence2})
in a simple way (see Fig. (\ref{fig:schema}) upper panel). 
Just after the particle is deblocked from a pinning site,
one assumes that the applied force $f_a(i)$ decreases linearly,
$f_a(i+1)=f_a(i)-c$ (with $c>0$) with increasing $i$ till the
particle gets blocked again. Thus, as opposed to
the horizontal plateaus between two successive records in the I.I.D
model (as in Fig. (\ref{fig:sequence2})), the force profile in the linear-trend model
has a linear behavior with a negative slope, as shown schematically
in Fig. (\ref{fig:schema}) upper panel. The physical
rationale behind the decrease of the applied force is 
the dissipation during the avalanche motion 
(in particular we have a linear decrease if $f_a(i)$ is the 
elastic force between the particle and a drive that moves at 
a constant slow rate). In this model, at the end of an avalanche when
the particle gets stuck at a new site, the
force profile $f_a(i)$ jumps again to the corresponding record value of 
the landscape at that site and the process continues. The sequence of 
the force values at the beigining of an avalanche coincides with
the record series $\{R_1,\, R_2,\, R_3,\,\cdots\}$ of the landscape
(see Fig. (\ref{fig:schema}) upper panel).

Interestingly, it turns out that this linear-trend
model was also introduced originally in the statistics literature
by Ballerini and Resnick~\cite{BR85,BR87}, and has since been studied
extensively with numerous applications~\cite{B99,FWK10,WFK11,FWK12}.
The analysis of the linear-trend model with $c>0$ shows that the average
number of records grows linearly with $N$ for large $N$, 
i.e., $\langle M_N\rangle\approx a(c)\, N$ where the prefactor
$a(c)$ is nontrivial and nonuniversal, i.e., depends
on $f(x)$~\cite{BR85,FWK10,WFK11}. The record value distribution $q_k(R)$ (or
equivalently the distribution of the applied forces at the begining of the $k$-th 
avalanche) does approach a stationary distribution as $k\to \infty$ (as in
realistic systems) that depends on the tail of $f(x)$. Similarly, the
avalanche size distribution $\pi_k(n)$ also approaches a stationary
distribution $\pi(n)$ as $k\to \infty$, however this stationary distribution
$\pi(n)$ does not have a power-law tail (rather an exponential
cut-off) for large $n$ as expected in real systems. 
In summary, the linear-trend model does reproduce some features
of avalanches in realistic depinning systems, but not all.

In this paper, we introduce a simple modification of the linear-trend 
model, which we call the $c$-record model. The model is defined more 
precisely in the next section where we also provide a 
summary of our main results. This model allows exact 
solutions for the three observables $\langle M_N\rangle$, $q_k(R)$ and 
$\pi_k(n)$. We show that these observables as well as the correlation 
structure between records and their ages reproduce the features observed 
in realistic systems.

\section{\label{sec:record_process} $c$-record model: 
the definition and a summary of main results}

\noindent{\bf The model.} The $c$-record model that we study in this paper is defined as follows.
Once again, we consider an infinite I.I.D landscape $\{x_1,\, x_2,\, x_3,\cdots\}$
of quenched pinning forces defined on a $1$-d lattice, where
each entry is chosen independently from a continuous PDF $f(x)$. 
Let $\{R_1,\, R_2,\, R_3,\cdots \}$
denote the record-series of this landscape. As in the simple I.I.D model, the particle
starts from site $i=1$ and it can leave the this site when the local
applied force $f_a$ exceeds the pinnning force $x_1$. The only difference
in our model from that of the linear-trend model discussed in the previous 
section, is how the force profile $f_a(i)$ behaves between two avalanches.
In the linear-trend model, the force profile following a record decreases linearly till
the next record (as in Fig. (\ref{fig:schema}) upper panel).
In contrast, in the $c$-record model, we assume that following a record, the force
decreases by $c$ only in the first step, but after that it stays flat till it encounters the
next record (see Fig. (\ref{fig:schema}) lower panel).
More precisely, between two
successive record values $R_k$ at $t_k$ and $R_{k+1}$ at $t_{k+1}$
we now have
\begin{equation}
f_a(i)= R_k-c,\, {\rm for}\,\, i=t_k+1, t_k+2,\cdots, t_{k+1}-1\, . 
\label{fa1.c_record}
\end{equation}
At $i=t_{k+1}$, the force $f_a(i)$ undergoes a jump to the associated record value $R_{k+1}$
of the landscape.
The physical rationale behind this new model is that the 
dissipation in the force profile 
that occurs just after depinning is short-ranged in time, e.g., 
occurs only during the first hopping but stays constant afterwards.

The formation of records in this $c$-record depinning model can
be alternatively phrased in the language of standard time-series
discussed in the introduction. Consider, as before, an infinite I.I.D sequence
of entries $\{x_1,\,x_2,\,x_3\,\ldots\}$ each drawn from $f(x)$.
Here a record series $\{R_1,\,R_2,\, R_3,\ldots\}$ is
formed recursively, in the presence of a single parameter $c>0$, as follows.
If a record occurs at some step with record value $R$, a subsequent
entry will be a record only if its value exceeds $(R-c)\,\theta(R-c)$.  
Clearly, for
$c=0$, this is the standard record model discussed in the introduction. 
For $c>0$, this is precisely the $\delta$-{\em exceedence} model introduced
by Balakrishnan et. al.~\cite{Bala96} with $\delta=-c<0$.
As already discussed in the introduction, this model with $c>0$ was also
studied in Ref.~\cite{PSNK15} as the random adaptive walk (RAW) model
of biological evolution in a random fitness landscape.

In this paper, we have studied the three observables
$\langle M_N\rangle $, $q_k(R)$ and $\pi_k(n)$ in the $c$-record model,
both analytically
and numerically, for a class of $f(x)$'s. From the motivation
of the depinning pheomenon, our main interest is 
to determine if, when $k 
\to \infty$, the $c$-record process becomes `stationary' or not. We say 
that it is stationary if the distributions $q_k(R)$ and $\pi_k(n)$ have 
a well defined limit as $k \to \infty$, namely:
\begin{flalign}
\label{eqn:stat_condition}
    & \lim_{k \to \infty} q_k(R) = q(R) \\
    & \lim_{k \to \infty} \pi_k(n) = \pi(n) \nonumber
\end{flalign}
Below we summarize our main results.

\vskip 0.3cm

\noindent{\bf The summary of main results.} 
We find that for $c>0$, the behavior of all three observables depend
explicitly on $f(x)$ and $c$. For simplicity, we consider
positive random variables, i.e., $f(x)$ with a positive support.
In particular, three cases can be distinguished:

\begin{enumerate}
    
\item If $f(x)$ decays slower than any exponential function as $x\to \infty$, then 
the record process doesn't reach a stationary limit for any $c>0$, i.e.,
$q_k(R)$ and $\pi_k(n)$ do not have a $k$-independent limiting distribution
as $k\to \infty$. 
The average number of records $\langle M_N\rangle$ grows logarithmically with $N$ as in the 
standard record problem $c=0$ (but with a different subleading constant): 
    \begin{equation}
    \label{eqn:MN_logN_slower}
        \langle M \rangle_N = \ln N + O(1)
    \end{equation}

\item If $f(x)$ decays faster than any exponential function as $x\to \infty$
(this includes bounded distribution), 
then the average number of records grows linearly with $N$ at the leading order:
    \begin{equation}
    \label{eqn:MN_N_faster}
        \langle M \rangle_N = A_1(c) N + O(N)
    \end{equation}
where the amptitude $A_1(c)$ can be analytically determined in some cases, e.g., 
for uniform $f(x)$ over the interval $[0,1]$ (see Eq. (\ref{eqn:exp_MN_uniform})).
In this case,  
the record process also reaches a stationary limit for all $c>0$. We
compute the stationary record value distribution $q(R)$
and the stationary age distribution $\pi(n)$, analytically and numerically, 
in several examples of $f(R)$. 
In particular, for distributions with a finite support, we show
that $\pi(n)$ decays exponentially with $n$ for large $n$. Finally, for distributions
with unbounded support and $f(x) \to e^{-x^\gamma}$ as $x\to \infty$ with
$\gamma>1$, we show that $\pi(n)$ still has a power-law tail with an exponent
larger than $2$.

\end{enumerate}

The case $f(x)=\exp(-x)$ that separates these two behaviors
turns out to be marginal, with a striking phase transition 
at $c=1$. While for $0\le c\le 1$, there is no limiting stationary distribution
for $q_k(R)$ and $\pi_k(n)$ as $k\to \infty$, we show that for $c>1$, they
do approach stationary distributions. In particular, we find
that for $c>1$
\begin{eqnarray}
\label{eqn:exp_qR}
\label{qR_pi_stat_exp}
     &  q(R) = \lambda(c) e^{-\lambda(c)\, R}  \\
     & \nonumber \\
     & \pi(n\to \infty) = \frac{\lambda(c)(1-\lambda(c))}{n^{1+\lambda(c)}}
\end{eqnarray}
where $\lambda(c)$ is the unique positive root of the transcendental equation:
    \begin{equation}
    \label{eqn:c_vs_lmbd}
        c = -\frac{\ln(1-\lambda)}{\lambda}
    \end{equation}
The average number of records $\langle M_N\rangle$ is computed
exactly in Eq. (\ref{eqn:exp_MN}) and its large $N$ behavior also exhibits a
phase transition at $c=1$. we show that 
    \begin{equation}
           \label{eqn:recordsnumber}
        \langle M \rangle_N = \begin{cases}
        \frac{1}{1-c}\ln N - \mu(c) +O(\frac{1}{N})& 0\le c < 1 \\
        \\
        \ln^2 N + O(\ln N) & c=1 \\
        \\
        A_0(c) N^{\lambda(c)} + \frac{1}{1-c} \ln N +O(1) &  c > 1
        \end{cases}
    \end{equation}
where the exponent $\lambda(c)$ depends continuously on $c$ and is
given in Eq. (\ref{eqn:c_vs_lmbd}). Note that $\lambda(c)$ increases monotonically
with increasing $c\ge 1$: $\lambda(c)\to 0$ as $c\to 1^{+}$, while 
$\lambda(c)\to 1$ only when $c\to \infty$.  
The explicit expression of the constant $A_0(c)$ is given in equation 
(\ref{eqn:A0_solution}). The constant $\mu(c)$ can be evaluated using 
the method explained in appendix (\ref{app:exp_case_MN_asymp}).
We also provide careful numerical checks for all our analytical formulae. 

As mentioned in the introduction, precisely this marginal case
$f(x)=e^{-x}$ was also studied in Ref.~\cite{PSNK15}  
in the context of RAW model in evolutionary biology and indeed, this
striking phase transition at $c=1$ was already noticed there.
In Ref.~\cite{PSNK15}, two of the three observables namely $\langle M_N\rangle$
and $q_k(R)$ (but not $\pi_k(n)$) were studied in detail, but using 
different notations, language and method.
In order to compare our results to those of Ref. \cite{PSNK15}, it is
useful to provide a dictionary of notations for the reader.
In the limit of large genome size $L$, the RAW model 
studied in \cite{PSNK15} becomes equivalent, in the ensemble sense, 
to our $c$-record model with $N\sim L$.
In this limit, our average number of records
$\langle M_N\rangle$ for large $N$ then translates to the mean length
of the adaptive walk $D_{\rm RAW}(L\sim N)$ in \cite{PSNK15} for large $L$.
Furthermore, our record value distribution $q_k(R)$ is
precisely $Q_{l=k}(y=R,L\to \infty)$, the probability for the adaptive walker to take
$l=k$ steps to arrive at a local fitness $c(l-L)+y$ in the limit
of large genome size $L\to \infty$. Hence, to summarize, for large
$L$ (or equivalently for large $N$ in our notation)
\begin{eqnarray}
& & N\sim L\,; \quad \langle M_N\rangle\equiv D_{\rm RAW}(L\sim N)\,;  \nonumber \\
& & q_k(R)\equiv Q_{l=k}(y=R,L\to \infty)\, .
\label{dictionary}
\end{eqnarray}

With the precise translation in Eq. (\ref{dictionary}) we can now compare
our results with those of Ref.~\cite{PSNK15}. We start with the asymptotic
large $N$ behavior of the average number
of records $\langle M_N\rangle$.
For $c\le 1$, our leading order large $N$ results for $\langle M_N\rangle$ in 
Eq. (\ref{eqn:recordsnumber}) agree
with that of Ref.~\cite{PSNK15}, though the subleading constant $-\mu(c)$ for $0\le c<1$ was
not computed in \cite{PSNK15}. However, for $c>1$, our result in 
Eq. (\ref{eqn:recordsnumber}) is much richer (characterized by a power law growth
of $\langle M_N\rangle$ with an exponent depending continuously on the
parameter $c$) and different from
that of \cite{PSNK15} where $\langle M_N\rangle \sim O(N)$ was reported
(see Eq. (15) of \cite{PSNK15}). 
We find that the growth of $\langle M_N\rangle$
with increasing $N$ becomes linear only when $c\to 1$, since only in
that limit $\lambda(c)\to 1$ in Eq. (\ref{eqn:recordsnumber}).

Next we turn to $q_k(R)$ for $c>1$. In this case, an exact summation 
formula for $q_k(R)$ was derived for all $c$ in \cite{PSNK15} (see their 
Eq. (8)). In the limit $k\to \infty$, this sum is convergent only for 
$c>1$. In Ref. \cite{PSNK15}, this sum was not analysed in the limit 
$k\to \infty$, since they didn't need it in their problem. 
In fact, it can be checked 
that their Eq. (8), in the limit $k\to \infty$ and for $c>1$,
satisfies the fixed point differential equation,
$q'(R)= q(R+c)-q(R)$ (see later in Eq. (\ref{eqn:exp_qR_stationary_diff_eq})),
whose solution is precisely a single pure exponential
$q(R)= \lambda(c)\, e^{-\lambda(c)\, R}$, with $\lambda(c)$
given by the positive root of Eq. (\ref{eqn:c_vs_lmbd}). This
is a rather nontrivial confirmation that two
different methods lead to the same solution.
 
\begin{figure}
\centering
\includegraphics[width=\linewidth]{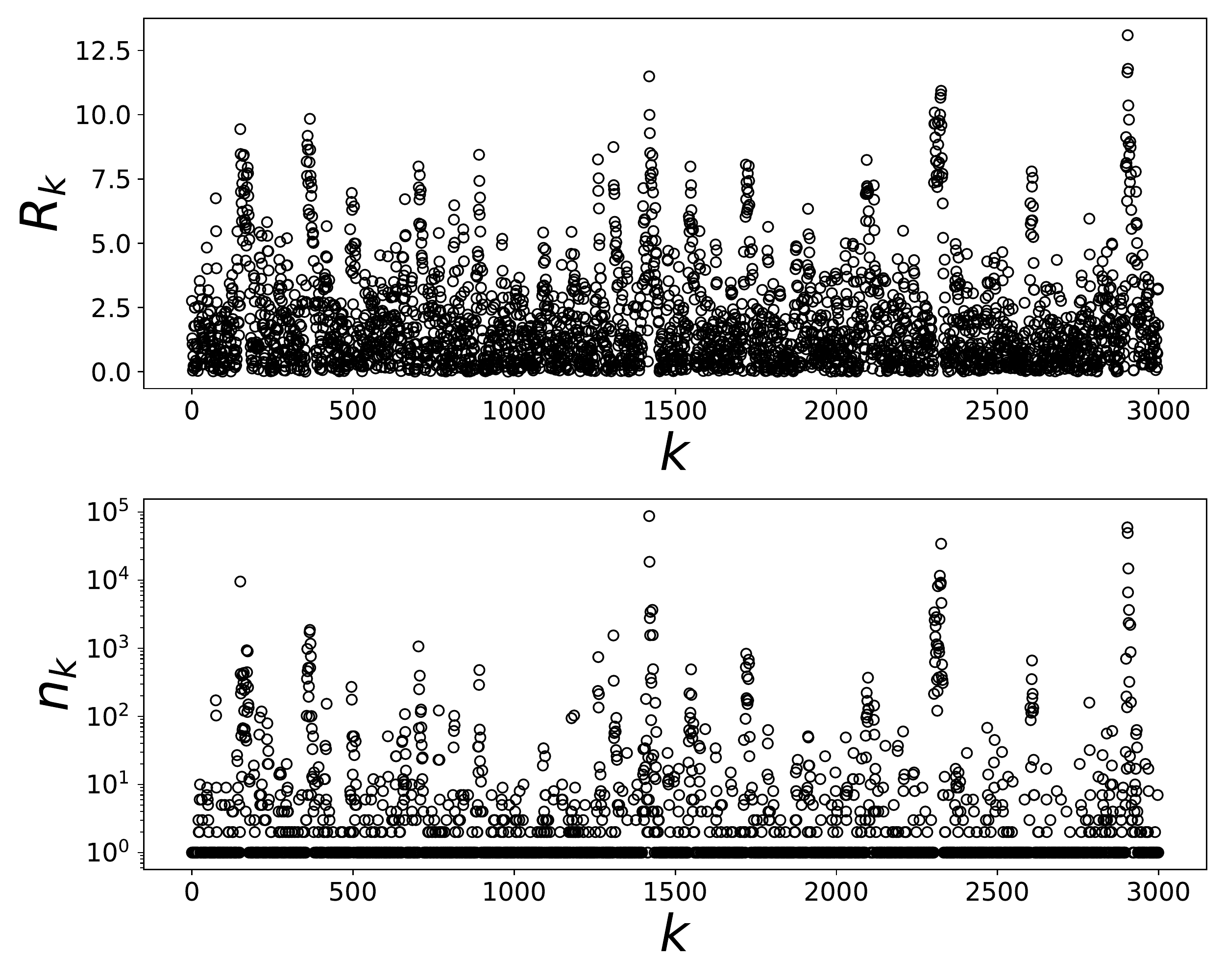}
\caption{Typical sequence of $c$-record values (upper panel) and 
their ages (lower panel) computed from a series of exponential number and 
using $c=1.5$. Note that large record values and large ages are organized in 
well defined clusters.}
\label{fig:exp_records_sequence}
\end{figure}

Let us finally conclude this summary section by mentioning that
we have also studied the correlations between record values
in the stationary state when it exists.
In that case, the sequence of records display
remarkable clustering properties 
(see Fig.~\ref{fig:exp_records_sequence}  upper panel)): 
after a large record value, we observe other large record values 
followed by a swarm of smaller record values. 
In the exponential case $f(x)=e^{-x}$, 
we show that the correlation between
record values $\langle R_k R_{k+\tau}\rangle$ decreases exponentially with
increasing $\tau$. The value of $c$ controls the correlation
length: when $c \to 1$ from above, the correlation lengths diverges as $\sim
1/(c-1)^2$ reminiscent of the critical phenomena (see Fig.~\ref{fig:corr_exp})).
We also study the clustering property of record ages that correspond to 
avalanches (see Fig.~\ref{fig:exp_records_sequence} lower panel)).
This property is qualitatively similar to what is observed in seismic 
catalogs.

\vskip 0.3cm

\vskip 0.3cm

\section{Setting up the recursion relations for the three observables}
\label{recursion}

In this section, we show how to set up the basic recursion relations
to compute the three observables in the $c$-record model: (i)
the mean number of records $\langle M_N\rangle$ up to $N$ steps,
(ii) the distribution $q_k(R)$ of the value of the $k$-th record 
and (iii) the distribution $\pi_k(n)$ of the time interval $n_k$ between
the $k$-th and the $(k+1)$-th record. For simplicity, we will
assume throughout that we have an infinite series of I.I.D entries
$\{x_1,\, x_2,\,x_3,\,\cdots\}$, each drawn from a continuous
$f(x)$ which has a positive support, i.e., the entries are
non-negative random variables. 

\subsection{\label{sec:recursive_rels_MN} The Number of Records }
In this subsection we derive the exact recursion relation that we have 
used to compute $\langle M \rangle_N$. It turns out that the
main object that we need for this computation is the
joint probability density $P_N(M,R_M=R)$ 
that in the first block of size $N$ of the infinite series of random 
entries there are exactly $M$ records and that the last record has value $R$.
This quantity $P_N(M,R)$ satisfies a closed recursion relation
\begin{multline}
    \label{eqn:recurrence_PMNR}
    P_N(M,R) = P_{N-1}(M,R) \theta(R-c) \int_{0} ^{R-c} f(x) dx  + \\
    +f(R) \int_0^{R+c} P_{N-1}(M-1,R') dR'
\end{multline}
with $\theta(x)$ the Heaviside step function. It is straightforward to 
understand Eq. (\ref{eqn:recurrence_PMNR}): the first term on the right hand side (r.h.s) 
accounts for the event when the $N$-th entry is not a record, while
the second term corresponds to the event when the $N$-th entry
is a record. In the first case, given that the last record has value $R$,
the value of the $N$-th entry
$x_N$ must be less than $R-c$ which happens with probability
$\int_{0} ^{R-c} f(x) dx$. In the second case, the $N$-th entry
is a record with value $R$, hence the previous record $R'$
must be less than $(R+c)$ explaining the second term on the r.h.s
in Eq. (\ref{eqn:recurrence_PMNR}). The recursion relation
(\ref{eqn:recurrence_PMNR}) starts from the initial condition
\begin{equation}
P_1(M,R)= \delta_{M,1}\, f(R)\, ,
\label{in_cond.1}
\end{equation}
since the first entry, by convention, is a record.
The recursion relation (\ref{eqn:recurrence_PMNR}), starting
from (\ref{in_cond.1}), is nontrivial to solve since 
it relates $P_N(M,R)$ at $R$ to its integral up to $R+c$ at step $N-1$, 
making it a non-local integral equation for any $c>0$.  

In order to compute $\langle M\rangle_N$, namely 
the average number of records as a function of $N$, we introduce 
$\tilde{Q}(R,z,s)$ as the double generating function of $P_N(M,R)$:
\begin{equation}
\label{eqn:q_tilde_definition}
    \tilde{Q}(R,z,s)=\sum_{N=1}^\infty \sum_{M=1}^\infty P_N(M,R) z^N s^M
\end{equation}
Using equation (\ref{eqn:recurrence_PMNR}) with the initial condition 
(\ref{in_cond.1}), we obtain the following equation 
for (\ref{eqn:q_tilde_definition}):
\begin{multline}
\label{eqn:recurrence_QR}
    \tilde{Q}(R) \left[1-zF(R-c) \theta(R-c) \right]  = \\
    zs f(R) \left[1+ \int_0^{R+c} \tilde{Q}(R') dR'\right]
\end{multline}
For simplicity, we omitted the arguments $z$ and $s$ of $\tilde Q$. Evidently,
Eq. (\ref{eqn:recurrence_QR}) is also non-local with respect to 
$R$ for any $c>0$. 

Even though the full joint distribution $P_N(M,R)$ of the number of records $M$
and the last record value $R$ is of interest as it contains several
interesting informations, we will focus on the simplest quantity, namely
the mean number of records $\langle M \rangle_N$.
To compute this,  we need to extract $P_N(M)$, the 
distribution of the number of records $M$ up to $N$ steps. This is
obtained by integrating over the value $R$ of the last record up to $N$ steps
\begin{equation}
    \label{eqn:PNM_definition}
     P_N(M)=\int_0^\infty P_N(M,R) dR \, .
\end{equation}
The double generating function of $P_N(M)$ is then related to $\tilde{Q}(R,s,z)$
simply by
\begin{equation}
\label{eqn:PNM_definition_gf}
    \sum_{N=1}^\infty \sum_{M=1}^\infty P_N(M) z^N s^M = 
\int_0^\infty \tilde{Q}(R,z,s) dR \, 
\end{equation}
where $\tilde{Q}(R,z,s)$ is the solution of the integral 
equation (\ref{eqn:recurrence_QR}).
Using Eq. (\ref{eqn:PNM_definition_gf}), the average number of 
records can be obtained from the relation
\begin{eqnarray}
\label{eqn:MN_by_derivatives}
\langle M \rangle_N & = & \sum_{M=1}^{\infty} M\, P_N(M) \nonumber \\
& = &\frac{1}{N!}\, 
\partial^N_z  \partial_s   \int_0^\infty \tilde{Q}(R,z,s) dR \,
|_{z=0,s=1}\, .
\end{eqnarray}
Solving the recursion relation (\ref{eqn:recurrence_QR}) for arbitrary $f(x)$
seems hard. However, we were able to compute $\tilde{Q}$ and from it
$\langle M_N\rangle$ explicitly for two special cases: the
exponential distribution $f(x)= e^{-x}\,\theta(x)$
and the uniform distribution 
$f(x)=\mathbb{I}_{[0,1]}(x)$ with $\mathbb{I}_{[a,b]}(x)$ denoting the indicator 
function which is $1$ if $x$ belongs to the interval $[a,b]$ and
zero otherwise.

\subsection{\label{sec:recursive_rels_qR} The record value 
distribution }
In this subsection we derive an exact recursion relation for the $k$-th record 
value distribution $q_k(R)$ in the $c$-record model.
We start from the joint 
conditional probability $q(R,n|R')$ that, 
given a record with value $R'$ has occurred at some instant in the
infinite sequence, 
the next record has value $R$ and
the age of the current record with value $R'$ is $n$, i.e, there are $n$ steps
separating the current record and the next record. This conditional
probability can be very simply computed
\begin{equation}
\label{eqn:Rn_cond_prop}
q(R,n|R')=\begin{cases}
f(R)\, \delta_{n,1}\, & R'<c \\ 
& \\
f(R)\, \left[F(R'-c)\right]^{n-1} \, \theta(R-R'+c)\, & R'>c
\end{cases}
\end{equation}
where we recall that $F(R)= \int_{0}^{R} f(y)\, dy$ is the cumulative
distribution of each entry of the underlying time-series. Eq. (\ref{eqn:Rn_cond_prop})
is easy to interpret: 
when the previous record $R'$ is smaller than $c$,
the very next entry with any positive value $R>0$ will be a record,
indicating that $n=1$ is the only possible value.
In contrast, when $R'>c$, assuming that there are exactly 
$n-1$ entries separating two 
successive records
with values $R'$ and $R$ (with $R>R'-c$), each of these intermediate entries
must be less that $R'-c$ (in order that none of them is a record), explaining
the factor $\left[F(R'-c)\right]^{n-1}$. The probability that the
$n$-th entry is a record is simply $f(R)\, \theta(R-R'+c)$.
By summing over all possible age values $n \geq 1$ we obtain the 
distribution of the next record value conditioned on the previous one:
\begin{eqnarray}
\label{eqn:R_cond_prop}
q(R|R') &= & \sum_{n=1}^{\infty} q(R,n|R') 
= f(R)\, \theta(c-R') \nonumber \\
&+& \frac{f(R)}{1-F(R'-c)}\, \theta(R-R'+c)\, \theta(R'-c) \nonumber \\
\end{eqnarray}
Equations (\ref{eqn:Rn_cond_prop}) and (\ref{eqn:R_cond_prop}) are 
particularly useful when simulating directly the $c$-record process (see 
appendix \ref{app:simulation}). 

A recursion relation for $q_k(R)$ can then be set up using this
conditional probbaility $q(R|R')$ as 
\begin{eqnarray}
\label{eqn:recurrence_qRk}
q_{k+1}(R)& = & \int_0^{\infty} q_k(R')\, q(R|R')\, dR' 
 = f(R) \int_0^{c} q_k(R') dR'  \nonumber \\ 
&+& f(R) \int_c^{R+c} \frac{q_k(R')}{1-F(R'-c)} dR'\, , \nonumber \\ 
\end{eqnarray}
with the initial condition $q_1(R)=f(R)$. This relation can be
easily understood as follows. The first term takes into account
the event when the record $R'$ is in the range $0\le R'\le c$. In
this case, the entry immediately after this record is a record
with probability $f(R)$. 
The second term
on the r.h.s in Eq. (\ref{eqn:recurrence_qRk}) accounts
for the contributions coming from the case when $c\le R'\le R+c$.
In this case, one needs to use $q(R|R')$ from Eq. (\ref{eqn:R_cond_prop})
and integrate over all allowed values of $R'$. 
This relation
(\ref{eqn:recurrence_qRk}) was also derived in \cite{PSNK15}
using different notations and for different observables, in the
context of the RAW model of evolutionary biology. Furthermore,
for $c=0$, this relation was used in Ref.~\cite{RP06} for studying 
the global temperature records.

When the stationary limit of $q_k(R)$ 
exists, $q(R)=\lim_{k \to \infty} q_k(R)$ has to satisfy the following 
fixed point equation
\begin{eqnarray}
\label{eqn:qR_sc}
q(R) & = & f(R)\, \int_0^c q(R') dR' \nonumber \\
    & + & f(R)\, \int_c^{R+c} \frac{q(R')}{1-F(R'-c)}\, dR' \, .   
\end{eqnarray}
This is also a non-local differential equation for any $c>0$ and is
hard to solve for general $f(x)$. Again, we will see later
that for $f(x)=e^{-x}\theta(x)$ (with $c>1$) and for the uniform distribution,
it is possible to obtain explicitly the fixed point stationary
solution of Eq. (\ref{eqn:qR_sc}).

\subsection{\label{sec:recursive_rels_pi} The age distribution of records }

In this subsection, we derive a recursion relation for $\pi_k(n)={\rm Prob.}(n_k=n)$ denoting
the distribution of the age $n_k$ of the $k$-th record, in an infinite I.I.D series.
The distribution $\pi_k(n)$ can be obtained
from the previously defined conditional probability $q(R,n|R')$ 
(\ref{eqn:Rn_cond_prop})
and the record value distribution $q_k(R)$ as follows.
\begin{eqnarray}
\label{eqn:pi_k_definition}
\pi_k(n) &= &\int_0^\infty \int_0^\infty q(R,n|R')\, q_k(R')\, 
dR\, dR' \nonumber  \\
& = & \delta_{n,1}\, \int_0^c q_k(R')\, dR' \nonumber \\ 
&+ & \int_c^\infty (1-F(R'-c)) F^{n-1}(R'-c) q_k(R') dR' \nonumber \\
\end{eqnarray}
where we used $F(R'-c)=0$ if $R'-c \leq 0$. The stationary 
distribution $\pi(n)$, when it exists, is obtained by taking the limit $\lim_{k \to 
\infty} \pi_k(n)$ and using $q(R)=\lim_{k \to \infty} q_k(R)$:
\begin{eqnarray}
\label{eqn:pi_stationary}
\pi(n) & = & \delta_{n,1}\, \int_0^c q(R')\, dR' \nonumber \\
    & +& \int_c^\infty (1-F(R'-c)) F^{n-1}(R'-c)\, q(R')\, dR' \, .  \nonumber \\
\end{eqnarray}
Thus, knowing the fixed point distribution $q(R)$ of the record value when it exists,
one can compute the stationary age distribution using Eq. (\ref{eqn:pi_stationary}).
Later, we will compute $\pi(n)$ explicitly for the two solvable cases, namely
the exponential distribution $f(x)=e^{-x}\theta(x)$ with $c>1$
and the case of the uniform distribution over $[0,1]$.

\section{\label{sec:exp_case}Exponential case}
In this section we study in detail the exponential case with $f(x)=\exp(-x)\,\theta(x)$. 

\subsection{\label{sec:exp_case_1}Number of records}
For the exponential distribution equation (\ref{eqn:recurrence_QR}) 
reduces to:
\begin{multline}
\label{eqn:QR_exp_case}
    \tilde{Q}(R) \left[1-z(1-e^{-(R-c)}) \theta(R-c) \right]  = \\
    z\,s\, e^{-R}\, \left[1+ \int_0^{R+c} \tilde{Q}(R') dR'\right] 
\end{multline}
The non-locality manifest in Eq. (\ref{eqn:recurrence_QR}) makes it hard to 
find its general solution. Remarkably, for the exponential $f(x)$, 
this is possible. Performing a rather involved calculation, 
reported in appendix 
(\ref{app:exp_case_PNM}), we computed the generating function for 
$P_N(M)$ defined in Eq. (\ref{eqn:PNM_definition_gf}). Its explicit 
form is given in Eq. (\ref{eqn:a0_exp_first_form}) from which we 
extracted the average number of records $\langle M \rangle_N$, obtaining
\begin{equation}
    \label{eqn:exp_MN}
    \langle M \rangle_N = N! \sum_{m=0}^{N-1} (-1)^m 
\frac{\prod_{k=1}^m (k-1+e^{-kc})}{(m+1)!^2 (N-m-1)!}
\end{equation}
This result for $\langle M_N\rangle$ is exact for all $N\ge 1$. 
For example, the first few values of $N$ yield
\begin{flalign}
\label{eqn:exp_MN_some_values}
    & \langle M \rangle_1 = 1 \\
    & \langle M \rangle_2 = 2 \left(1-\frac{1}{4}e^{-c}\right) \\
    & \langle M \rangle_3 = 3 \left[1-\frac{4}{9}e^{-c} + 
\frac{1}{18} e^{-3c}\right]
\end{flalign}
A non trivial check of equation (\ref{eqn:exp_MN}) is for $c=0$. By 
plugging $c=0$ the term $\prod_{k=1}^m (k-1+e^{-kc})$ simplifies to 
$k!$. Thus equation (\ref{eqn:exp_MN}) becomes:
\begin{equation}
\label{eqn:MN_at_c_0}
\langle M \rangle_N = \sum_{m=0}^{N-1} \binom{N}{m+1}  (-1)^m \frac{1}{(m+1)}
\end{equation}
The above expression can be shown to coincide with the well known 
result $\langle M \rangle = 1 + 1/2+1/3+\dots+1/N$. 
This can be done by considering the difference 
$\langle M \rangle_N - \langle M \rangle_{N-1}$.
While the exact formula (\ref{eqn:exp_MN}) is useful to compute the result for 
moderate values of $N$, the asymptotic behavior of $\langle M \rangle_N$ 
for large $N$ is difficult to extract from it. 

\begin{figure}
\includegraphics[width=\linewidth]{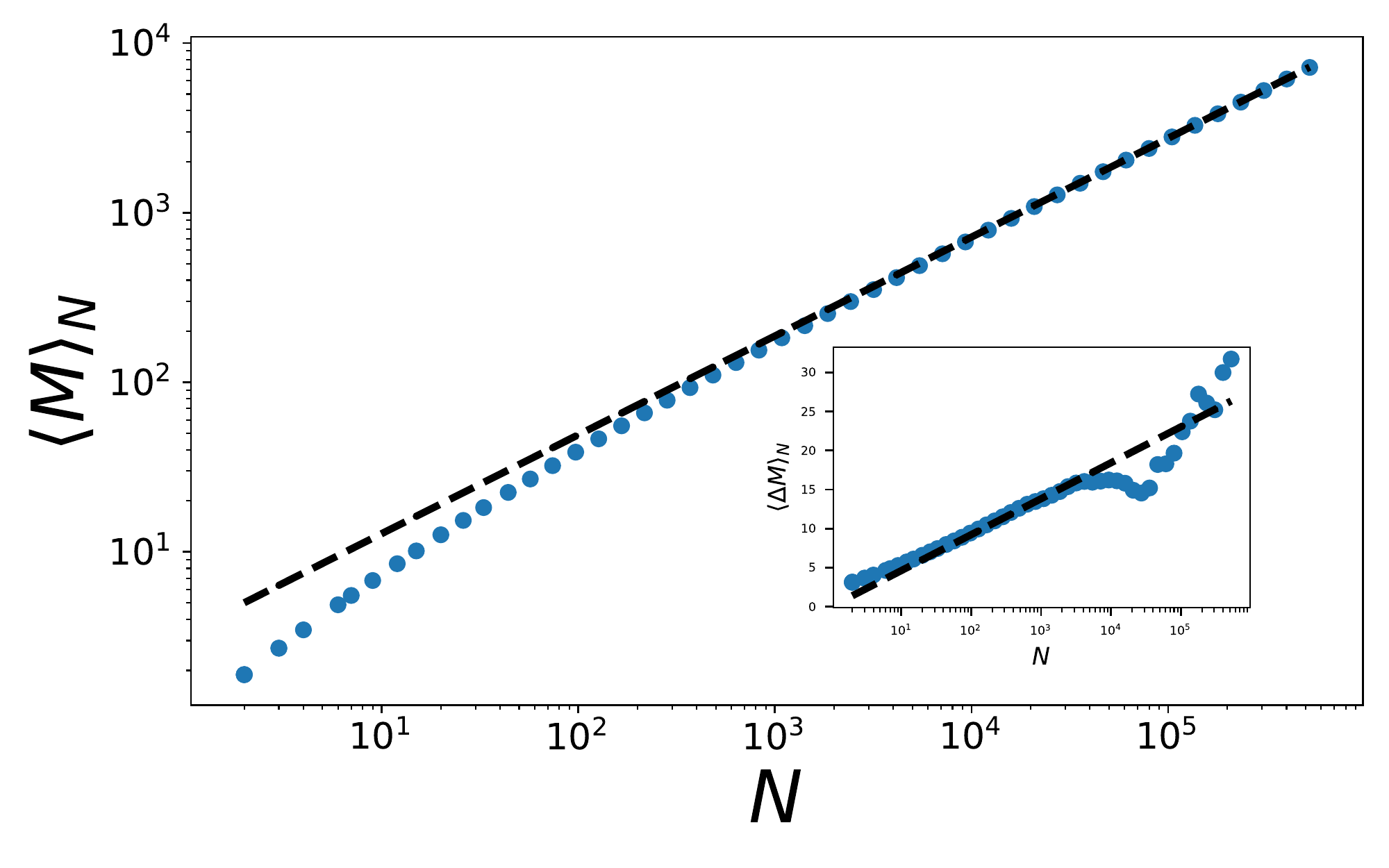}
\caption{
Exponential case: mean number of $c$-records for $c=1.5$. Blue symbols 
correspond to numerical data. Black dashed line corresponds to the 
analytical prediction $A_0(c)N^{\lambda(c)}$. The values of the 
constants are $A_0(1.5)=3.4376$, $\lambda(1.5)=0.5828$ (obtained from 
equations (\ref{eqn:c_vs_lmbd}) and (\ref{eqn:A0_solution}) using 
Mathematica). Inset: the subleading behavior $\langle \Delta M \rangle_N 
=A_0(c)N^{\lambda(c)}-\langle M \rangle_N$. Black dashed line 
corresponds to $-\frac{1}{1-c} \ln N$.  }
\label{fig:M_N_exp_supercrit}
\end{figure}

\begin{figure}
\includegraphics[width=\linewidth]{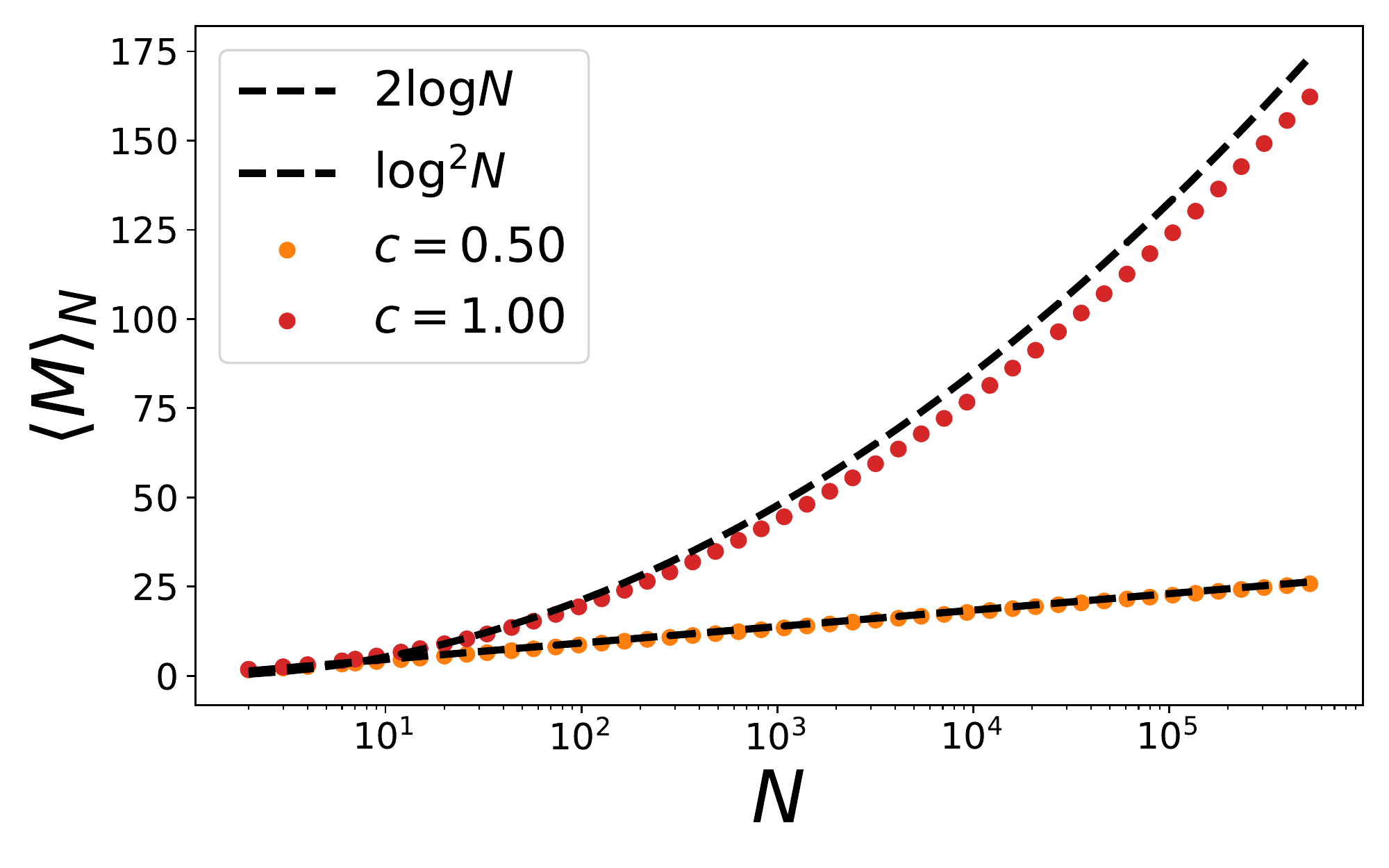} \caption{ 
Exponential case: mean number of $c$-records for $c=0.5$ (red circles) 
and for $c=1$ (orange circles). Circles are numerical data. Black dashed 
lines correspond to the analytical predictions of 
Eq.(\ref{eqn:recordsnumber}). For $c=0.5$ we find $\mu(c)=1.42878\ldots$ 
computed using Mathematica by the method of appendix 
(\ref{app:exp_case_MN_asymp}).}

\label{fig:M_N_exp_subcrit}
\end{figure}

Indeed, to extract the large $N$ behvaior, we use a different
approach reported in appendix (\ref{app:exp_case_MN_asymp}) 
which leads to our main result in Eq. (\ref{eqn:recordsnumber}) and shows the existence of a 
phase transition at $c=1$. The presence of the phase transition at $c=1$ in the
asymptotic behavior of $\langle M_N\rangle$ in Eq. (\ref{eqn:recordsnumber})
is also related to the fact that both $q_k(R)$ and $\pi_k(n)$
has stationary limiting distributions as $k\to \infty$, only for $c>1$
as shown in the next subsection \ref{sec:exp_case_2}. 
To check the validity 
of our analytical prediction we also performed direct numerical simulations 
using $N=10^6$ random variables and averaging over $10^4$ realizations. 
The results are shown in Fig. (\ref{fig:M_N_exp_supercrit}) for $c=1.5$ 
and in Fig. (\ref{fig:M_N_exp_subcrit}) for $c=0.5$, as well as for the 
marginal case $c=1$.

\subsection{\label{sec:exp_case_2}Stationary record and age statistics}

\begin{figure}
 \includegraphics[width=\linewidth]{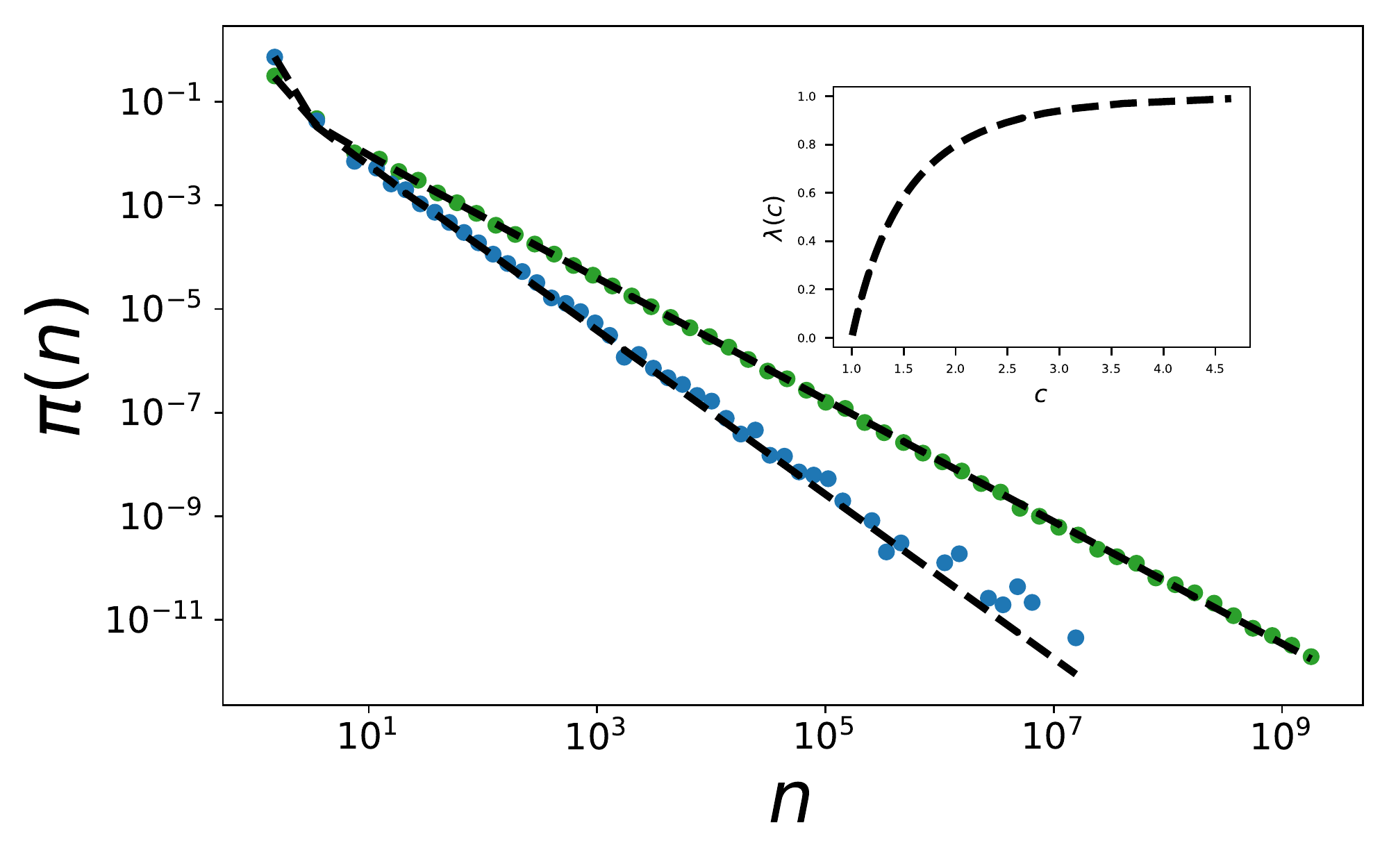}
\caption{
Exponential case: age statistics of the stationary $c$-records process 
for $c=1.5$ (blue circles) and for $c=1.1$ (green circles). Numerical 
data are averaged over $10^4$ realizations. Black dashed line 
corresponds to the analytical predictions of Eq.(\ref{eqn:pi_exp}). The 
inset shows $\lambda(c)$ as a function of $c$. }
\label{fig:gap_distr_exp}
\end{figure}

The solution of Eq. (\ref{eqn:recurrence_qRk}) for a given $k$ has 
been studied by Park et al. in \cite{PSNK15} for the case 
$f(x)=\exp(-x)\, \theta(x)$. Here we focus instead on the stationary limit and we 
seek the solution of Eq. (\ref{eqn:qR_sc}). By taking a derivative 
with respect to (w.r.t.) $R$ of equation (\ref{eqn:qR_sc}) we get
\begin{equation}
\label{eqn:qR_sc_derivative}
    q'(R)=\frac{f(R)}{1-F(R)} q(R+c) + \frac{f'(R)}{f(R)} q(R)
\end{equation}
For the exponential case, using $f(R)=\exp(-R)$ and 
$1-F(R)=\exp(-R)$, equation (\ref{eqn:qR_sc_derivative}) reduces to:
\begin{equation}
\label{eqn:exp_qR_stationary_diff_eq}
    q'(R)=q(R+c)-q(R)
\end{equation}
To solve this equation we use the ansatz:
\begin{equation}
    \label{eqn:exp_qR.0}
    q(R)=\lambda\, e^{-\lambda R}
\end{equation}
Equation (\ref{eqn:exp_qR_stationary_diff_eq}) becomes
\begin{equation}
    \label{eqn:exp_c_lmbd}
    1-\lambda=e^{-\lambda c}
\end{equation}
A positive solution for $\lambda=\lambda(c)$ exists only for $c>1$. This 
means that for $c \leq 1$ there is no stationary regime. This conclusion 
was already pointed out in \cite{PSNK15} based on scaling arguments and 
numerical simulations. Here we perform an explicit calculation which is 
also confirmed by the independent calculation of the large $N$ expansion 
of $\langle M \rangle_N$ report in the appendix 
(\ref{app:exp_case_MN_asymp}). The stationary average record value is 
$1/\lambda(c)$ (see inset of Fig. (\ref{fig:gap_distr_exp})) and has the 
following limiting behaviors:

\begin{equation}
\label{eqn:avg_R_exp}
    \langle R \rangle(c) = 1/\lambda(c) = \begin{cases}
    1/2(c-1) & c \to 1^+ \\
    1 & c \to \infty
    \end{cases}
\end{equation}
The divergence as $c \to 1^+$ is one of the fingerprints of the absence 
of a stationary state for $c \leq 1$. 

We now turn to the age distribution $\pi_k(n)$. As in the case of $q_k(R)$,
a stationary distribution $\pi(n)$ in the limit $k\to \infty$ exists
only for $c>1$. For $c\le 1$, $\pi_k(n)$ depends explicitly on $k$
even in the $k\to \infty$ limit. To derive the stationary
age distribution $\pi(n)$ for $c>1$, we substitute
$q(R)$ from Eq. (\ref{eqn:exp_qR.0}) into Eq. (\ref{eqn:pi_stationary}).
Upon carrying out the integration explicitly,
we obtain the following exact expression of $\pi(n)$ in the stationary 
state valid for all $n\ge 1$ and $c>1$
\begin{equation}
    \label{eqn:pi_exp}
    \pi(n) = \lambda \delta_{n,1} + (1-\lambda) 
\frac{\lambda \Gamma(n)\Gamma(\lambda+1)}{\Gamma(n+\lambda+1)}\, ,
\end{equation}
where $\Gamma(z)=\int_0^{\infty} e^{-x}\, x^{z-1}\, dx$ is the standard
Gamma function.
The stationary age (or avalanche size) distribution $\pi(n)$ is then the 
sum of two contributions:
a delta peak at $n=1$ and the  
Yule-Simon form for $n \geq 1$. 
Using the asymptotic behavior,
$\Gamma(n)\Gamma(\alpha)/\Gamma(n+\alpha) \to 1/n^\alpha$ for large $n$, 
we unveil a beautiful power law behavior of the stationary age 
distribution (see Fig.~(\ref{fig:gap_distr_exp}))
\begin{eqnarray}
\label{eqn:pi_exp_limit_inf}
    \pi(n \to \infty) = \frac{\lambda(1-\lambda)}{n^{1+\lambda}}
\end{eqnarray}
where $\lambda\equiv \lambda(c)$ is given by the positive root of Eq. (\ref{eqn:exp_c_lmbd})
for $c>1$.
Note that the power law exponent $(1+\lambda(c))$ 
continuously varies with $c$. When $c \to \infty$, 
the exponent $\lambda(c)\to 1$, thus
strengthening the amplitude of the delta peak and
in addition, for large $n$, the power law tail behaves as $\pi(n)\sim 1/n^2$.

We now show that the power law decay of $\pi(n)\sim n^{-(1+\lambda(c))}$ 
for large $n$ in Eq. (\ref{eqn:pi_exp_limit_inf}) for $c>1$
is completely consistent with the result $\langle M_N\rangle\sim N^{\lambda(c)}$
for large $N$ in the third line of Eq. (\ref{eqn:recordsnumber}).
To see this, we use a simple scaling argument. Given 
$\pi(n)\sim n^{-(1+\lambda(c))}$ for large $n$ and the length of the series $N$,
the mean inter-record distance
(mean age of a record) scales, for large $N$, as
\begin{equation}
\langle n\rangle= \sum_{n=1}^N n\, \pi(n)\sim N^{1-\lambda(c)}\, .
\label{mean_age.1}
\end{equation}
Consequently, the mean number of records, which is identical to the mean number of
inter-record intervals, up to step $N$ scales for large $N$ as
\begin{equation}
\langle M_N\rangle \sim \frac{N}{\langle n\rangle}\sim N^{\lambda(c)}\, ,
\label{meanr.1}
\end{equation}
which reproduces the third line of Eq. (\ref{eqn:recordsnumber}) for $c>1$.
In appendix (A), this result is proved more rigorously.

\begin{figure}
\includegraphics[width=\linewidth]{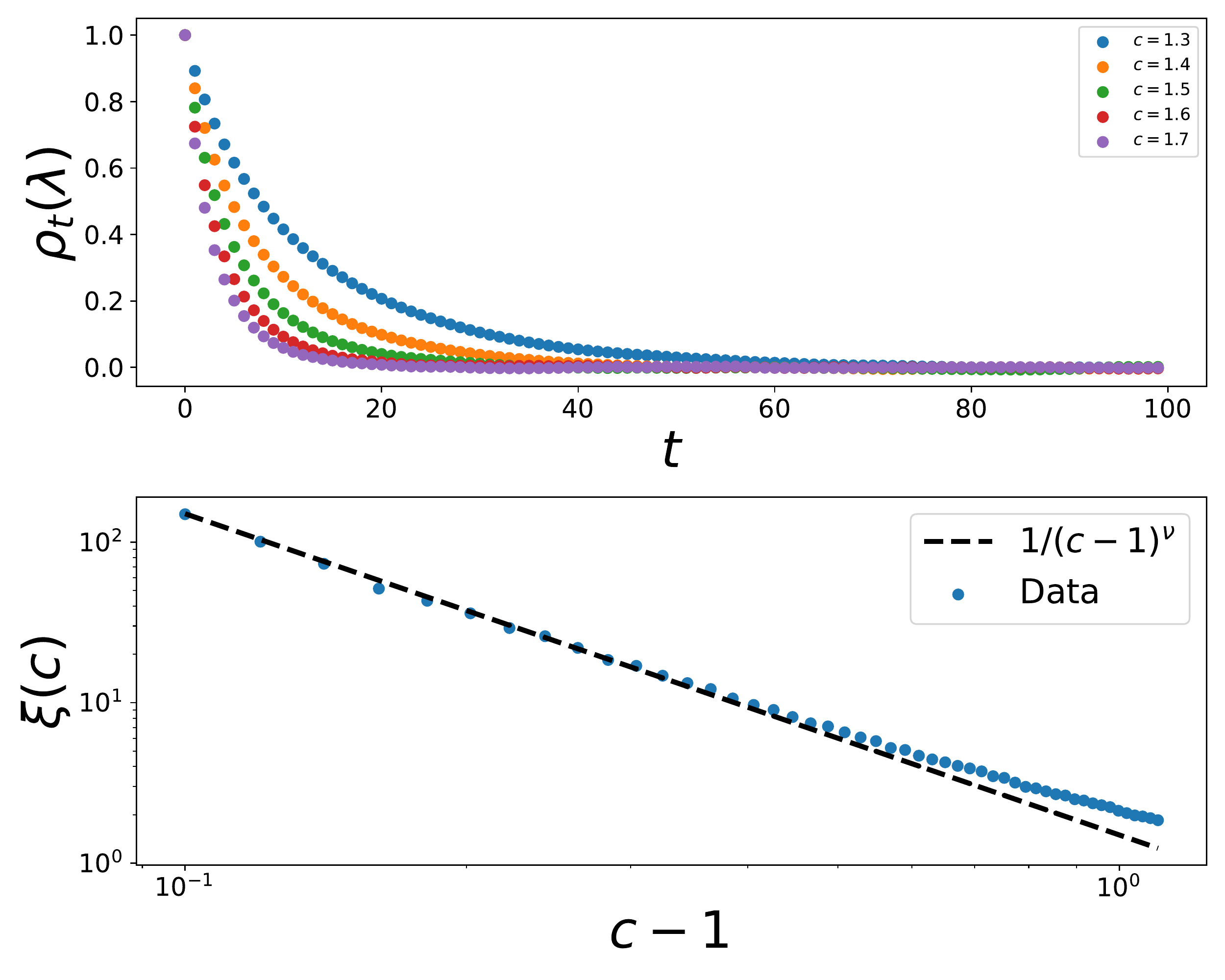}
\caption{Upper panel: record correlation function numerically computed as in 
equation (\ref{eqn:record_corr_func}). Lower panel: correlation length 
$\xi(c)$ as a function of $c-1$. We fitted an exponent $\nu \approx 2$. 
}
\label{fig:corr_exp}
\end{figure}

\subsection{\label{sec:exp_case_3}Record correlations}
One of the most interesting features of the $c$-records statistics is 
shown in Fig. (\ref{fig:exp_records_sequence}). We focus on a sequence 
of records when the stationary regime is already reached. The sequence 
tends to cluster in patterns where record values are high, followed by 
events of smaller value. The corresponding sequence of ages shows a 
similar behavior: when the record values are high we observe large ages, 
while the age is of the order $1$ when the records are small.\\ To 
understand this behavior we can first observe that in the classical 
record case $c=0$, even if there is no stationary regime, all the 
records values are strongly correlated. As a fingerprint of this, the 
sequence of record values is strictly increasing. For $c>0$ this is not 
always the case. For example we know that the $c$-record process of an 
exponential series with $c>1$ has a stationary state and the 
correlations have a finite range (which corresponds to the typical size 
of the correlated patterns in Fig.~(\ref{fig:exp_records_sequence})). 
To characterize this behaviour we focus on the record values and study 
the following correlation function:
\begin{equation}
\label{eqn:record_corr_func}
    \rho_c(\tau) \equiv \frac{\text{Cov}(R_k R_{k+\tau})}{\text{Var}(R_k)}  
\end{equation}
By definition $\rho_c(0)=1$. In appendix (\ref{app:corr_exp}) we compute 
the correlation between two successive records, namely $\rho_c(1)$ which 
reads
\begin{equation}
\label{eqn:record_corr_func_1}
    \rho_c(1) = (1-\lambda(c))(1-\ln(1-\lambda(c)))\, .
\end{equation}
When $c \to \infty$, from Eq.~(\ref{eqn:c_vs_lmbd}), $\lambda(c) 
\to 1$ and $\rho_c(1) \to 0$. This means that when $c \to \infty$ the 
records become uncorrelated. Indeed every random variable $\{ x_1, x_2, 
\dots \}$ becomes a record. On the other hand as $c \to 1$, $\lambda(c) 
\to 0$ and $\rho_c(1) \to 1$. This result signals that the correlations 
are very strong and we study numerically how fast they decay for the 
exponential case. In Fig. (\ref{fig:corr_exp}) we compute the 
correlation function (\ref{eqn:record_corr_func}) of a stationary record 
sequence for the exponential distribution and for different values of 
$c$. As in standard critical phenomena, the correlation length diverges 
when one approaches the critical point $c=1$. The numerical curves in 
Fig.~(\ref{fig:corr_exp}) are well fitted by an exponential law as 
$\rho_c(\tau) = \exp(-\tau/\xi(c))$.  Using this fit we estimate the 
correlation length and find that it diverges as,
$\xi(c) \sim 1/(c-1)^\nu$, with $\nu 
\approx 2$ (see Fig.~(\ref{fig:corr_exp})). This result is coherent with 
the estimation of $\xi(c)$ via $\xi(c) \sim -1/\ln \rho_c(1)$ coming 
from $\exp(-1/\xi(c))=\rho_c(1)$. Indeed as $c \to 1$, $\lambda(c) \sim 
2(c-1)$ and $-1/\ln \rho_c(1) \sim (c-1)^{-2} $.

\section{Stretched Exponential $f(x)$}
\label{stretched}

Let us recall from the summary of results in Section \ref{sec:record_process}
that a stationary limiting distribution for $q_k(R)$ and $\pi_k(n)$ exist for any $c>0$, if
$f(x)$ decays faster than $e^{-x}$ for large $x$. In the
complementary case when $f(x)$ has a slower than exponential tail, there is no stationary
limiting distribution. In the borderline case
$f(x)=e^{-x}$ one has a phase transition at $c=1$, separating the non-stationary phase ($c\le 1$)
and the stationary phase $(c>1)$, as demonstrated in detail in the previous section.
In this section, we will investigate the two complementary cases of $f(x)$: respectively
with a `faster' and a `slower' than exponential tail. We will do so by choosing $f(x)$ from
the stretched exponential family, defined on the positive real axis $x\ge 0$,
\begin{equation}
\label{stretched_f.1}
    f_{\rm stretched}(x)=\frac{\gamma}{\Gamma
\left(\frac{1}{\gamma}\right)}\, e^{-x^\gamma}\, \quad {\rm for}\,\, \gamma>0
\end{equation}
with cumulative $F_{\text{stretched}}(x)=1-
\frac{\Gamma\left(\frac{1}{\gamma},
x^\gamma\right)}{\Gamma \left(\frac{1}{\gamma}\right) }$, where
$\Gamma(s,t)=\int_t^\infty x^{s-1} e^{-x} dx $ the incomplete gamma
function. This family of $f(x)$ in Eq. (\ref{stretched_f.1}) includes the `borderline' exponential as the special
case $\gamma=1$. Furthermore, the case $\gamma>1$ would correspond to a `faster' 
than exponential, while $\gamma<1$ would correspond to `slower' than exponential tail. 

The reason why $\gamma=1$ is a borderline case, i.e., why the presence
of a finite $c>0$ afftects the record statistics differently for $\gamma<1$
and $\gamma>1$ can be understood intuitively using the extreme value 
statistics as follows. Consider first $c=0$. Let us consider the
first $N$ steps of the infinite I.I.D sequence. The value of the
last record in this series of size $N$ then coincides, for $c=0$,
with the global maximum $X_{\max}$ up to $N$ steps. For the stretched
exponential $f(x)$ in Eq. (\ref{stretched_f.1}), it is well known
from the theory of extreme value statistics (for a recent review
see~\cite{MPS20})
that while the mean value $\langle X_{\max}\rangle \sim (\ln N)^{1/\gamma}$
for large $N$, the variance scales as 
\begin{equation}
\sigma^2=\langle X_{\max}^2\rangle-\langle X_{\max}\rangle^2
\sim (\ln N)^{2(1-\gamma)/\gamma}\, .
\label{var_M}
\end{equation}
Hence, for $0<\gamma<1$, the width of the fluctuation grows with
increasing $N$, while for $\gamma>1$ it decreases for large $N$.
Now, imagine switching on a small $c>0$. If $0<\gamma<1$, an addition
of a finite offset $c$ will not affect the record statistics, since
it is much smaller than the fluctuation of the record value for large $N$.
In contrast, for $\gamma>1$ where the width is of $O(1)$ for large $N$,
the record value and its statistics will obviously be more sensitive to 
a finite offset $c$. The case $\gamma=1$
is thus marginal. In fact, this change of behavior at $\gamma=1$ for the
stretched exponential family was also noticed in the asymptotic
behavior of the density of near extreme events, i.e., the density
of entries near the global maximum in an I.I.D series~\cite{SM2007}. 
Below, we will
provide a more precise derivation of this change of behavior in
the record statistics at $\gamma=1$ due to a nonzero $c>0$. 

\begin{figure}
    \centering
    \includegraphics [width=\linewidth] {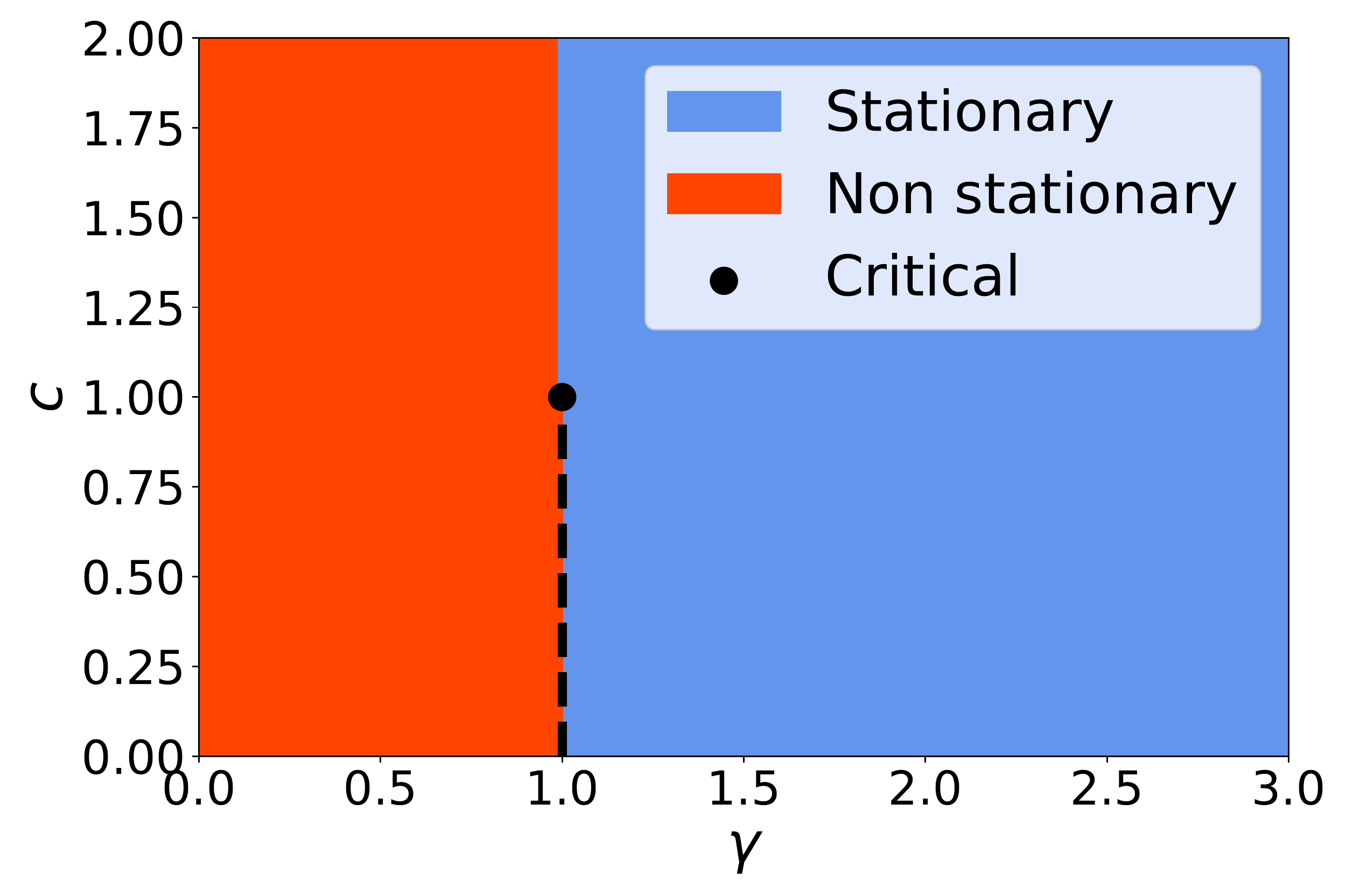}
    \caption{Phase diagram for the stretched exponential case. For 
$\gamma < 1$ no stationary limit exists for any $c>0$ while for $\gamma 
> 1$ it always exists. The exponential distribution $\gamma=1$
corresponds to the marginal case, for which 
a stationary state exists only for $c>1$.  }
\label{fig:phase_diagram}
\end{figure}

In this section, for simplicity we focus only on one observable, namely the record value distribution
$q_k(R)$. Our main goal here is to understand the criteria for having a stationary
distribution for $q_k(R)$ in the limit $k\to \infty$.
The other two observables $\langle M_N\rangle$ and $\pi_k(n)$ can also be studied
in principle, but we will skip them here to keep the paper shorter.
For the $c$-record model with
$f(x)$ belonging to this class, we then have two parameters $(\gamma,c)$.
Our goal is to find in which region in the $(\gamma\ge 0,\, c\ge 0)$ quadrant, we
have a stationary solution for $q_k(R)\to q(R)$ as $k\to \infty$. We will show
that this leads to an interesting phase diagram shown in Fig.~(\ref{fig:phase_diagram}).

We start from the general recursion relation for $q_k(R)$ in Eq. (\ref{eqn:recurrence_qRk}), valid for 
general $f(x)$.
We assume that there exists a stationary solution $q(R)$ which would then satisfy
the integral equation (\ref{eqn:qR_sc}). Our strategy would be to find, for $f(x)$
given in Eq. (\ref{stretched_f.1}), if Eq. (\ref{eqn:qR_sc}) allows 
a normalizable solution $q(R)$. If it does not, there is no stationary solution.
For later analysis, it is first convenient to define
\begin{equation}
q(R)= f(R)\, G(R)
\label{def_GR}
\end{equation}
and rewrite Eq. (\ref{eqn:qR_sc}) as
\begin{equation}
 G(R)=\int_0^c G(R')\, f(R')\, dR' + \int_c^{R+c} 
\frac{G(R') f(R')}{1-F(R'-c)} dR'
\label{GR.1}
\end{equation}
where we recall $F(R)=\int_0^R f(y)dy$.
By taking a derivative with respect to $R$ one gets
\begin{equation}
\label{eqn:GR_diff_eq}
G'(R)=\frac{f(R+c)}{1-F(R)}\, G(R+c) \, .
\end{equation}
This is a first-order non-local differential equation. We need only one `boundary' 
condition, i.e., the value of $G(R)$ at some point $R$, to fix
the solution uniquely. To find such a condition, we note that 
the solution of this differential 
equation need, in addition, to satisfy the original integral equation
(\ref{GR.1}). Substituting, e.g.,  $R=0$ in Eq. (\ref{GR.1}) gives a
condition
\begin{equation}
G(0)= \int_0^c G(R')\, f(R')\, dR' \, .
\label{G0.1}
\end{equation}
This shows that $G(0)$ is a constant and the solution must
satisfy this condition (\ref{G0.1}) self-consistently.
Another compatibility condition follows by investigating
the large $R$ limit of Eq. (\ref{GR.1}). If the limit $G(\infty)$
exists, one obtains
\begin{equation}
G(\infty)= \int_0^c G(R')\, f(R')\, dR' + \int_c^{\infty}
\frac{G(R') f(R')}{1-F(R'-c)} dR' \, .
\label{GR.2}
\end{equation}
For arbitrary $f(x)$, 
it is hard to find a general solution to Eq. (\ref{eqn:GR_diff_eq}) with
boundary condition (\ref{G0.1}) or (\ref{GR.2}).
Hence, below we focus on the stretched exponential class in Eq. (\ref{stretched_f.1}),
for which the
first-order equation (\ref{eqn:GR_diff_eq}) reduces to
\begin{equation}
\label{eqn:GR_diff_eq_sexp}
G'(R)= \frac{e^{-(R+c)^\gamma}}{\left[\int_R^\infty dR' e^{-{R'}^\gamma}\right]}\, G(R+c)\, .
\end{equation}
Note that for $\gamma=1$, Eq. (\ref{eqn:GR_diff_eq_sexp}) reduces to
$G'(R)= e^{-c}\, G(R+c)$ and this leads to the nontrivial solution
$q(R)= \lambda(c)\, e^{-\lambda(c)\, R}$ for all $R\ge 0$ with
$\lambda(c)$ given in Eq. (\ref{eqn:exp_c_lmbd}), as 
discussed in the previous section.

\vskip 0.3cm

\noindent {\em The case $c=0$ and arbitrary $\gamma>0$.} Let us first start with $c=0$ case with arbitrary $\gamma>0$. In this case, Eq. (\ref{eqn:GR_diff_eq_sexp})
becomes local in $R$ whose general solution can be easily found
\begin{equation}
G(R)= G(R_0)\, \frac{\int_{R_0}^{\infty}\,e^{-x^{\gamma}}\, dx}{\int_{R}^{\infty}\, e^{-x^{\gamma}}\, dx}\, ,
\label{GR_sol.1}
\end{equation}
where $R_0$ is arbitrary.
The condition (\ref{G0.1}) says that for $c=0$, $G(0)=0$ identically.
If we choose $R_0=0$ in Eq. (\ref{GR_sol.1}), then using $G(0)=0$, we see
that the only possible solution for $G(R)$ is just $G(R)=0$ for all $R$.
Consequently, using Eq. (\ref{def_GR}), we get $Q(R)=0$ which
evidently can not be normalized to unity.
This indicates that there is no stationary solution $q(R)$ for $c=0$
and arbitrary $\gamma>0$.
In other words, there is no stationary solution on the horizontal axis $c=0$ in the phase diagram
in Fig. (\ref{fig:phase_diagram}).

\vskip 0.3cm

\noindent {\em The case $0<\gamma<1$ and arbitrary $c> 0$.} 
In this case, we want to show that there is no limiting stationary solution
$q(R)$ (see the (red) shaded region in the phase diagram in Fig. (\ref{fig:phase_diagram})).
In other words, we will show that a solution to Eq. (\ref{eqn:GR_diff_eq}) satisfying
the condition (\ref{GR.2}) does not exist for $\gamma<1$ with
$c\ge 0$ arbitrary.
To show this, it is sufficient to investigate Eq. (\ref{eqn:GR_diff_eq}) for large
$R$, keeping $c\ge 0$ fixed. For large $R$, let us first assume
that the limit $G(\infty)$ in Eq. (\ref{GR.2}) exists.
Since $G(R)$ approaches a constant as $R\to \infty$,
it follows that for any arbitrary $c\ge 0$ and large $R$, we must have $G(R+c)\to G(\infty)$ on
the r.h.s of Eq. (\ref{eqn:GR_diff_eq}). This gives, for large $R$,
\begin{equation}
G'(R) \approx G(\infty)\, \frac{e^{-(R+c)^\gamma}}{\left[\int_R^\infty
dR'\, e^{-{R'}^\gamma}\right]} \, .
\label{GR_left.1}
\end{equation}
Now consider first the large $R$ behavior of the denominator on the r.h.s
of Eq. (\ref{GR_left.1}). It is easy to show that, to leading order for large $R$,
\begin{equation}
\int_R^{\infty} e^{-{R'}^\gamma}\, dR' \approx \frac{1}{\gamma}\, R^{\gamma-1}\, e^{-R^{\gamma}}\, . 
\label{GR_left.2}
\end{equation}
Substituting this in Eq. (\ref{GR_left.1}) one obtains for large $R$ and for any $\gamma>0$
\begin{eqnarray}
G'(R) &\approx & G(\infty)\, \gamma\, R^{\gamma-1}\, e^{-(R+c)^{\gamma}+ R^{\gamma}} \nonumber \\ 
&\approx&  G(\infty)\,\gamma\, R^{\gamma-1}\, e^{-\gamma\, c\, R^{\gamma-1}}\, .
\label{GR_larger.1}
\end{eqnarray}
Consider now the case $0<\gamma<1$. In this case, the argument of the exponential
on the r.h.s of Eq. (\ref{GR_larger.1}) vanishes, i.e., $e^{-\gamma\, c\, R^{\gamma-1}}\to 1$
as $R\to \infty$. Consequently, integrating Eq. (\ref{GR_larger.1}), one finds
that $G(R)\sim R^{\gamma}$ actually grows with increasing $R$ for $\gamma<1$. But this
is incompatible with the condition (\ref{GR.2}) and our starting assumption $G(\infty)$
is finite. In fact, this is also incompatible with the original
integral equation (\ref{GR.1}).
As $R\to \infty$, the l.h.s of Eq. (\ref{GR.1}) 
grows as $R^{\gamma}$, while the r.h.s approaches a constant since the
integral on the r.h.s is convergent as $R\to \infty$.
Hence we conclude that for $0<\gamma<1$ and $c\ge 0$, there is no stationary solution
for $G(R)$, and equivalently for $q(R)$ leading to the (red) shaded area
of the phase diagram in Fig. (\ref{fig:phase_diagram}).

\vskip 0.3cm

\noindent {\em The case $\gamma>1$ and arbitrary $c> 0$.}
Let us first check that in this case there is no obvious incompatibility
between the large $R$ behavior in Eq. (\ref{GR_larger.1}) and
the original integral equation (\ref{GR.1}). Indeed, for $\gamma>1$,
the exponential factor $e^{-\gamma\, c\, R^{\gamma-1}}$
on the r.h.s of Eq. (\ref{GR_larger.1}) decays rapidly, and integrating
over $R$ we find that $G(R)$ approaches a constant as $R\to \infty$,
which is perfectly compatible with Eq. ({\ref{GR.1}) in the large $R$ limit, 
or equivalently with
Eq. (\ref{GR.2}).
This already indicates that there is a normalizable stationary solution for any $c>0$
and $\gamma>1$. To compute explicitly this solution for arbitrary $c>0$ (and $\gamma>1$) 
still seems rather hard. Below, we compute this stationary distribution for $\gamma>1$ in two 
opposite limits: (i) $c\to 0^+$ and (ii) $c\to \infty$ limit.

\vskip 0.3cm

\noindent{\em The limit $c\to 0^+$ and $\gamma>1$ arbitrary.} We start with the 
$c\to 0$ limit with fixed $\gamma>1$. We fix $R$ in 
Eq. (\ref{eqn:GR_diff_eq_sexp}) and take the limit $c\to 0^+$.
To leading order for small $c$, we can approximate $G(R+c)\approx G(R)$
to make Eq. (\ref{eqn:GR_diff_eq_sexp}) local and also
expand $(R+c)^{\gamma}\approx R^{\gamma}+ \gamma\, c\, R^{\gamma-1}$
up to $O(c)$ with $R$ fixed. Eq. ({\ref{eqn:GR_diff_eq_sexp}) then
reduces to
\begin{equation}
\frac{1}{G(R)}\, \frac{d G(R)}{dR}\approx 
\frac{e^{-R^{\gamma}-c\,\gamma\, R^{\gamma-1}}
}{\left[\int_R^{\infty} dR'\, e^{-{R'}^{\gamma}}\right]}\equiv g_c(R)\, .
\label{GR_smallc.1}
\end{equation}
Note that only $c$-dependence appears through the factor 
$c\,gamma\, {R'}^{\gamma-1}$
inside the exponential in $g_c(R)$. For $\gamma>1$, this term contributes significantly
for large $R$, even when $c\to 0^+$. Hence, we can not neglect this $c$-dependent
term, especially for large $R$. One can then easily integrate Eq. (\ref{GR_smallc.1})
and obtain the solution as
\begin{equation} 
\label{eqn:GR_stretched_exp_small_c_solution}
G(R) \approx G(0)\,  \exp \left[\int_0^{R} g_c(x)\, dx\right]\, ,
\end{equation}
where $G(0)$ is a constant and $g_c(x)$ is defined in Eq. (\ref{GR_smallc.1}).
It is easy to show that for large $x$, $g_c(x)$ behaves as
$g_c(x)\approx \gamma\, x^{\gamma-1}\, e^{-\gamma\, c\, x^{\gamma-1}}$.
Hence for $\gamma>1$, the integral $\int_0^\infty g_c(x)\, dx$ is perfectly
convergent and is just a constant. Consequently, we find from 
Eq. (\ref{eqn:GR_stretched_exp_small_c_solution}) that 
$G(R)\to G(\infty)=G(0)\,\exp\left[\int_0^\infty g_c(x)\, dx\right]$ 
as $R\to \infty$. Thus the stationary solution
$q(R)$ in this $c\to 0^+$ limit is given by
\begin{equation}
q(R)\approx \frac{\gamma}{\Gamma
\left(\frac{1}{\gamma}\right)}\, e^{-R^{\gamma}}\, G(R)
\approx f(R)\, G(R)
\label{qr_smallc.1}
\end{equation}
where the function $G(R)$, given in 
Eq. (\ref{eqn:GR_stretched_exp_small_c_solution}), 
is well defined for any $R$ as long as $c\to 0^+$ (nonzero) and $\gamma>1$.
Thus, in the $c\to 0$ limit,
the stationary record value distribution 
$q(R)$ gets modified considerably from the parent distribution $f(R)$
by the multiplicative factor $G(R)$.

\vskip 0.3cm

\noindent{\em The limit $c\to \infty$ and $\gamma>1$ arbitrary.}
Next we consider the opposite limit 
$c \to \infty$. When $c \to \infty$ every random variable in the series
$\{x_1,\,x_2,\,x_3,\,\cdots\}$ is a record, hence 
$q(R) = f(R)$ and $G(R)=1$ for all $R\ge 0$. 
Now consider $c$ large, but not strictly infinite. 
In this case, the
function $G(R)$ will change from its flat value $G(R)=1$ that is valid
strictly for $c\to \infty$. However, we expect that $G(R)$, in the limit
$R\to \infty$, is not very sensitive to $c$, i.e., $G(\infty)=1$
even for finite but large $c$. However, for finite but fixed $R$,
we expect that $G(R)$ will deviate from its flat value $1$. 
To find this change in $G(R)$ for fixed $R$, to leading order for large
$c$, we can use the approximation 
$G(R+c)\approx 1$ on the r.h.s of Eq. (\ref{eqn:GR_diff_eq_sexp})
and solve the resulting first-order local equation, leading to
the solution
\begin{equation}
\label{GR_sol.2}
G(R) = 1 - \int_R^\infty \frac{e^{-(R'+c)^\gamma}}{\left[\int_{R'}^\infty
e^{-x^\gamma} dx\right]}\, dR'\, ,
\end{equation}
where we used the expected boundary condition $G(\infty)=1$ mentioned
above. For fixed $R$ and large $c$, we can approximate
$(R+c)^{\gamma}\approx c^{\gamma}+\gamma\, c^{\gamma-1}\, R$.
Using this approximation in the numerator of the integrand
in Eq. (\ref{GR_sol.2}) and rescaling $z= \gamma\, c^{\gamma-1}\, R'$,
we find that for $\gamma>1$ and in the scaling limit $c\to \infty$, $R\to 0$
such that the product $c^{\gamma-1}\, R$ is fixed
\begin{equation}
G(R)\approx 1- \frac{e^{-c^\gamma}}{c^{\gamma-1}\, \Gamma(\frac{1}{\gamma})}\,
e^{-\gamma c^{\gamma-1}\,R}\, .
\label{GR_sol.3}
\end{equation}
This is clearly compatible with the starting ansatz that $G(\infty)=1$.
Finally, using this in Eq. (\ref{def_GR}) we get the stationary solution
for $\gamma>1$ and large $c$ limit
\begin{equation}
q(R)\approx 
f(R)\, \left[1- \frac{e^{-c^\gamma}}{c^{\gamma-1}\, \Gamma(\frac{1}{\gamma})}\,
e^{-\gamma c^{\gamma-1}\,R}\right]\, .
\label{qR_sol.3}
\end{equation}
Thus, in the $c\to \infty$ limit, $q(R)$ approaches $f(R)$ with 
a small additive corection term as given in Eq. (\ref{qR_sol.3}).

Summarizing, for $\gamma>1$ in the phase diagram in Fig. 
(\ref{fig:phase_diagram}), the record value distribution
becomes stationary for large $k$, for any $c>0$.
In the two limits $c\to 0$ and $c\to \infty$, 
the stationary distribution $q(R)$ is given respectively
in Eqs. (\ref{qr_smallc.1}) and (\ref{qR_sol.3}). We have
checked these results numerically in Fig. (\ref{fig:sexp_qR})
for $\gamma=2$. In this figure, we see that as $c$ increases,
$q(R)$ progressively approaches $f(R)$.

\begin{figure}
   \includegraphics[width=\linewidth]{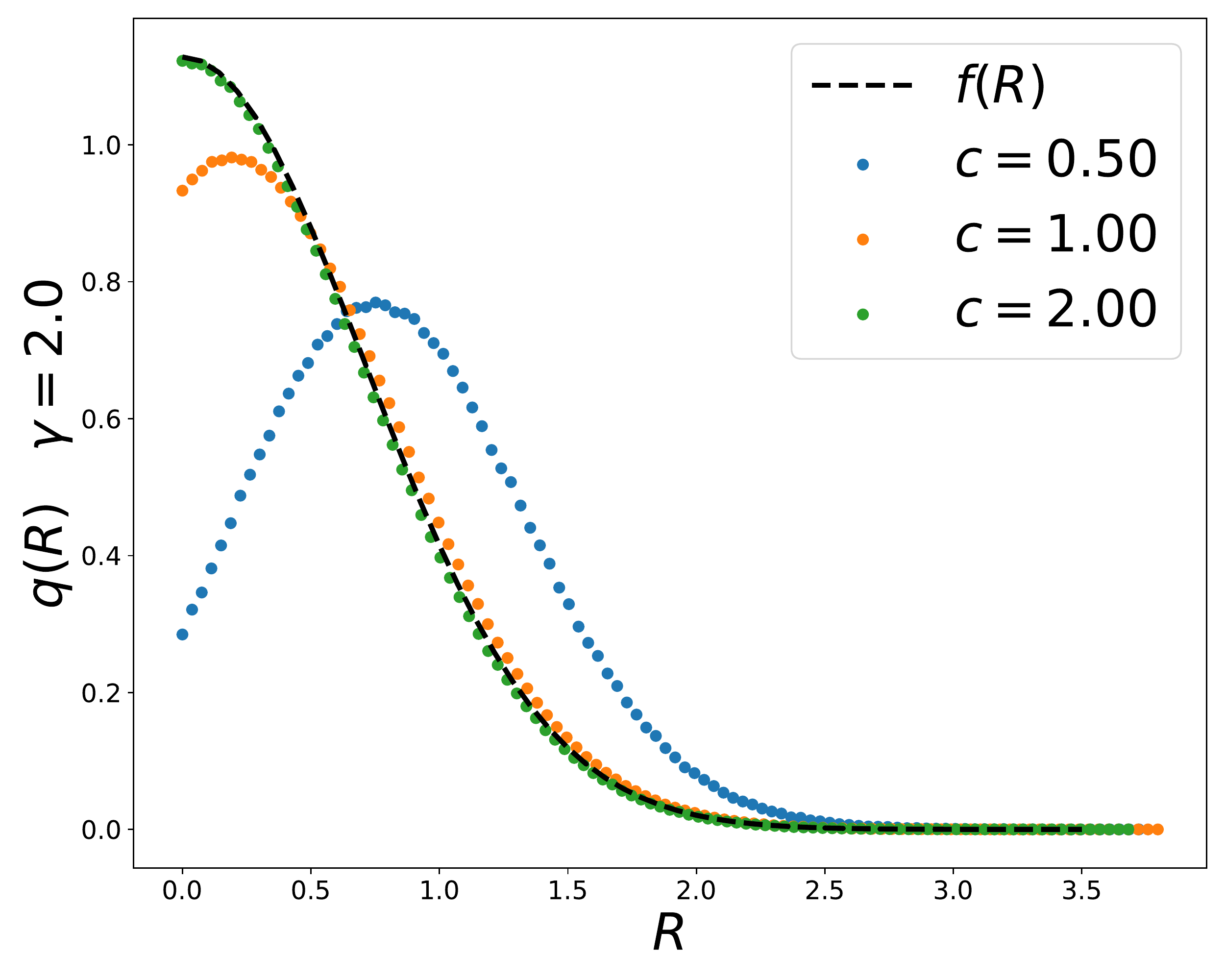}
\caption{Stretched exponential case: 
stationary record distributions $q(R)$ at $\gamma=2$ and 
for different $c$. As $c$ gets bigger, $q(R)$ approaches $f(R)$.}
\label{fig:sexp_qR}
\end{figure}

We have also computed numerically the mean number of records
$\langle M_N\rangle$ up to $N$ steps. As indicated in
the phase diagram in Fig.~(\ref{fig:phase_diagram}), we expect
that for $0<\gamma<1$ the record process is non-stationary, i.e.,
the effect of $c$ is insignificant and the system behaves similar to
$c=0$. Hence in this regime ((red) shaded region in the phase diagram
in Fig.~(\ref{fig:phase_diagram}), we expect $\langle M_N\rangle \simeq \ln N$
for large $N$, for any $c>0$. This prediction is verified numerically
in Fig. (\ref{fig:M_N_sexp_log}), for $\gamma=1/2$ and for 
several values of $c$. 
In contrast, for $\gamma>1$ and $c>0$ ((blue) shaded region
in the phase diagram in Fig.~(\ref{fig:phase_diagram})), 
we have a stationary phase.
In this case,
we expect a linear growth $\langle M_N\rangle \simeq A_1(c)\, N$ for
large $N$, with an $c$-dependent amplitude $A_1(c)\le 1$.
In the limit $c\to \infty$, we expect $A_1(c)\to 1$, since every entry
becomes a record in this limit. In Fig. (\ref{fig:M_N_sexp_lin}),
we verify this prediction for fixed $\gamma=2$ and for several values of $c$.

\begin{figure}
   \includegraphics[width=\linewidth]{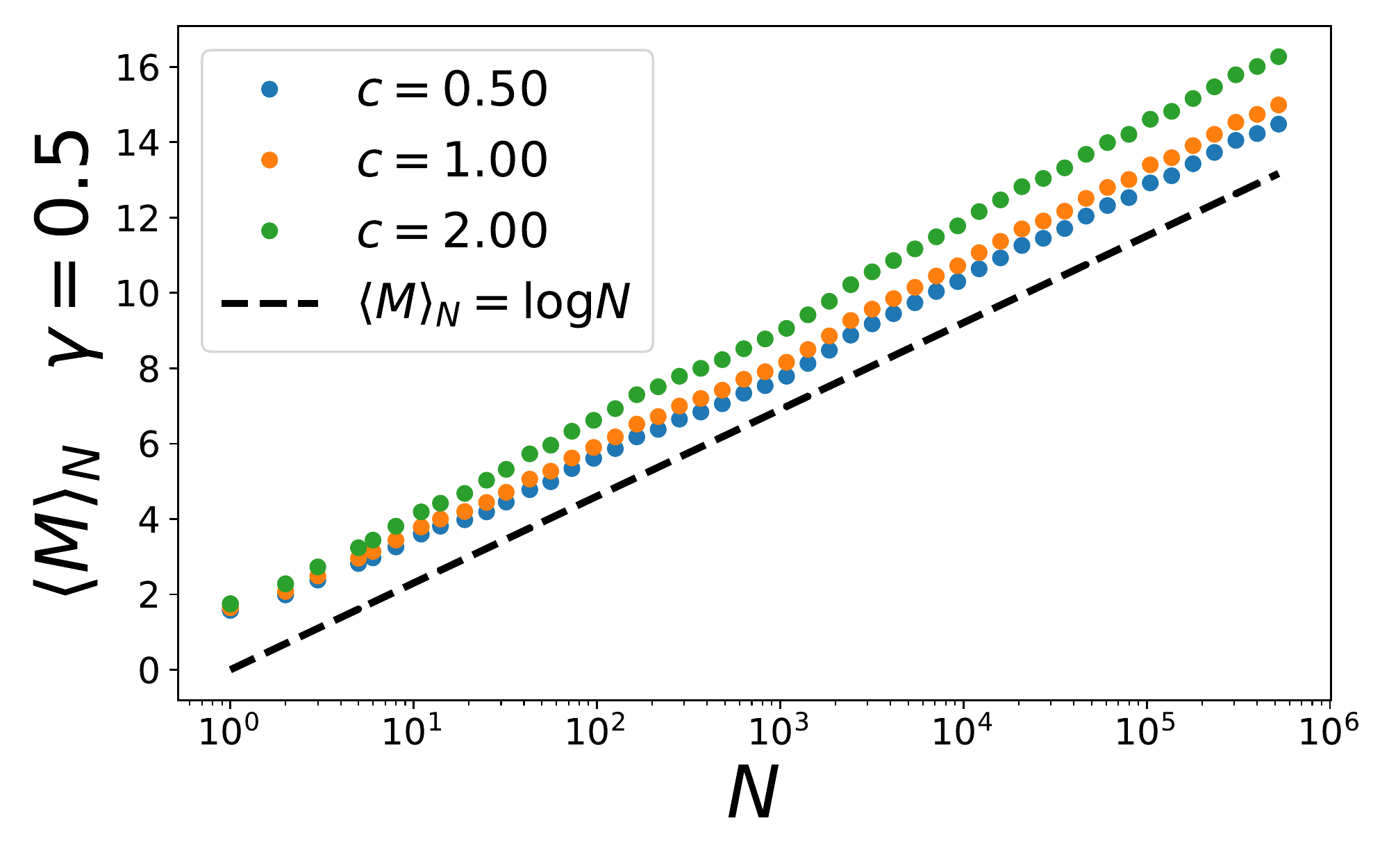}
\caption{Stretched exponential case: logarithmic growth of the mean number of $c$-records for different $c$ and $\gamma=0.50$. The dashed line is a guide to the eye.}
\label{fig:M_N_sexp_log}
\end{figure}

\begin{figure}[h!]
   \includegraphics[width=\linewidth]{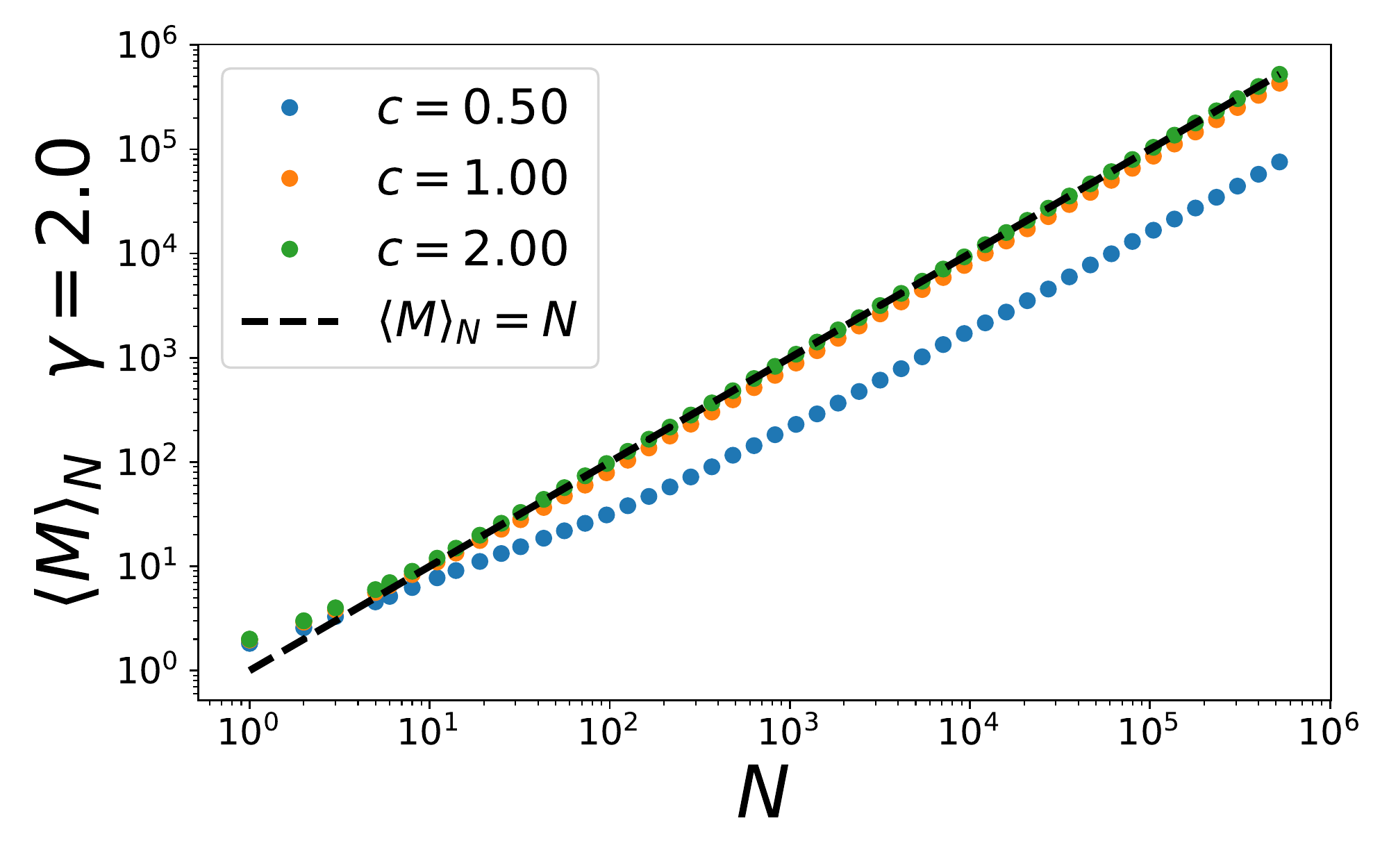}
\caption{Stretched exponential case: linear growth of the 
mean number of $c$-records for different $c$ and $\gamma=2$.}
\label{fig:M_N_sexp_lin}
\end{figure}

\section{Other distributions}
\label{other_dist}

In this section, we study other classes of $f(x)$. First we consider 
the uniform distribution of $f(x)$ over the bounded interval $[0,1]$ for which
we present exact analytical results for all three observables 
$\langle M_N\rangle$,
$q_k(R)$ and $\pi_k(n)$. We then consider more general bounded distributions.
Bounded distributions belong to the family of $f(x)$ with a `faster than'
exponential tail, hence we anticipate and demonstrate below that 
both $q_k(R)$ and $\pi_k(n)$
allow stationary limiting distributions as $k\to \infty$ for bounded $f(x)$. 
We then consider another
class of unbounded distribution, which we call the Weibull class
\begin{equation}
f_{\text{Weibull}}(x) = \gamma\, x^{\gamma-1}\, e^{-x^\gamma}\, 
\quad {\rm for}\,\, \gamma>0
\label{weibull.1}
\end{equation}
with $F_{Weibull}(x)=1-e^{-x^\gamma}$.
It turns out that this $f(x)$ is easier to sample numerically 
using the inverse Transform method, as explained in the appendix
(\ref{app:simulation}). We present detailed numerical results for all three
observables $\langle M_N\rangle$, $q_k(R)$ and $\pi_k(n)$ for this case.

\subsection{\label{sec:bounded_case}Bounded distributions}

The $c$-record process associated to I.I.D time series drawn from 
bounded distributions has a well defined stationary limit for 
any $c>0$. For simpicity we restrict the interval to the segment 
$[0,1]$. This implies that for any $c > 1$, every entry $x_i$ of the time 
series is a record.

\subsection{\label{sec:bounded_case_1} Uniform distribution }

We first consider the uniform distribution $f(x)=\mathbb{I}_{[0,1]}(x)$ 
and show that the mean number of records, the stationary record 
distribution and their age distribution can be explicitly computed for 
$1/2\le c \le 1$. For $0<c < 1/2$. the calculations are more cumbersome and we rely 
on Monte Carlo simulations.

To computed the mean number of records we study equation 
(\ref{eqn:recurrence_QR}) for the uniform distribution. 
Its expression simplifies to:
\begin{multline}
\label{eqn:QR_unif}
    \tilde{Q}(R) \left[1-z(R-c) \theta(R-c) \right]  =\\
    =z\,s+  z\,s\, \int_0^{\min(1,R+c)} \tilde{Q}(R') dR' 
\end{multline}
This equation is solved in appendix (\ref{app:bounded_case_PNM}) for 
$1/2\le c \le 1$
and the exact mean number of records reads:
\begin{equation}
\label{eqn:exp_MN_uniform}
   \langle M \rangle_N =  (2-c+\ln(c))N + \frac{1-c}{c} + \ln(c)
\end{equation}
Note that when $c=1$, we get $ \langle M \rangle_N =N$ as expected. In 
Fig. (\ref{fig:M_N_unif}) we 
numerically computed $\langle M \rangle_N$ for different values of 
$c$ and included the analytical predictions for $1/2\le c \le 1/2$. 

The stationary record distribution $q(R)$ satisfies Eq. 
(\ref{eqn:qR_sc_derivative}) together with the condition that $q(R)=0$ 
for $R>1$:
\begin{eqnarray}
\label{eqn:qR_unif_diff_eq}
     q'(R) = 
\begin{cases}
    0 &\quad \; 1-c<R<1 \\
    &\\
    \frac{q(R+c)}{1-R}  & \quad \; 0< R<1-c
\end{cases}
\end{eqnarray}
Eq. (\ref{eqn:qR_unif_diff_eq}) is valid for any $0 \leq c \leq 1$. 
For $c \geq 1/2$, it can be simply solved. In particular, imposing the 
global normalization, one gets:
\begin{equation}
\label{eqn:qR_unif_distr}
    q(R) = \frac{1}{2-c+\ln(c)} \begin{cases} 
    1 & 1-c<R<1 \\
    1-\ln \frac{1-R}{c} & 0 < R < 1-c
    \end{cases}
\end{equation}
In Fig.~(\ref{fig:uniform_case}) (upper panel) we show 
$q(R)$ for different values of $c$ along with the analytical predictions 
for $1/2\le c \le 1$.
 
The stationary age distribution $\pi(n)$ (for $1/2\le c \le 1$) 
follows from Eqs. (\ref{eqn:pi_stationary}) and (\ref{eqn:qR_unif_distr}):
\begin{multline}
\label{eqn:pn_unif_distr}
    \pi(n)=\frac{1+\ln(c)}{2+\ln(c)-c} \delta_{n,1} + \\
    +\frac{1}{2+\ln(c)-c} \frac{(1-c)^n}{n} \left[1-\frac{n(1-c)}{n+1}\right]
\end{multline}
$\pi(n)$ shows a delta peak at $n=1$, as in the exponential case. For 
large $n$, $\pi(n)$ decays exponentially over a 
characteristic scale $-1/\ln(1-c)$ that gets smaller as $c \to 1$. 
The fact that $\pi(n)$ has a 
well defined first moment is compatible with the scaling of the average 
number of records, namely $\langle M \rangle_N \propto N$ as $N \to 
\infty$.

\begin{figure}
   \includegraphics[width=\linewidth]{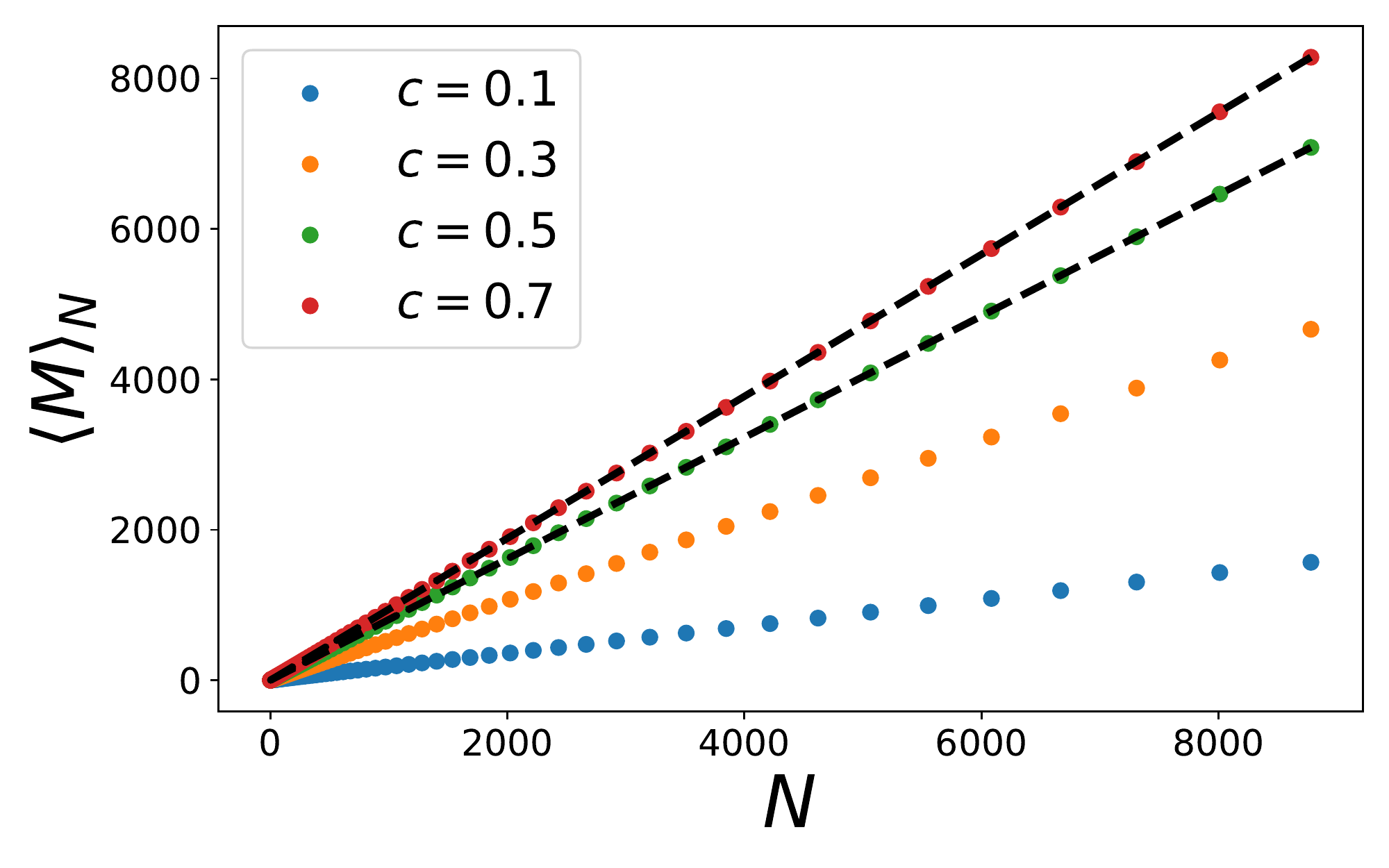}
\caption{Uniform case: mean number of $c$-records for different $c$. 
For $c \geq 1/2$ we include the analytical predictions 
in Eq. (\ref{eqn:exp_MN_uniform}) as shown by dashed lines.
The agreement is excellent.}
\label{fig:M_N_unif}
\end{figure}

\begin{figure}[h!]
   \includegraphics[width=\linewidth]{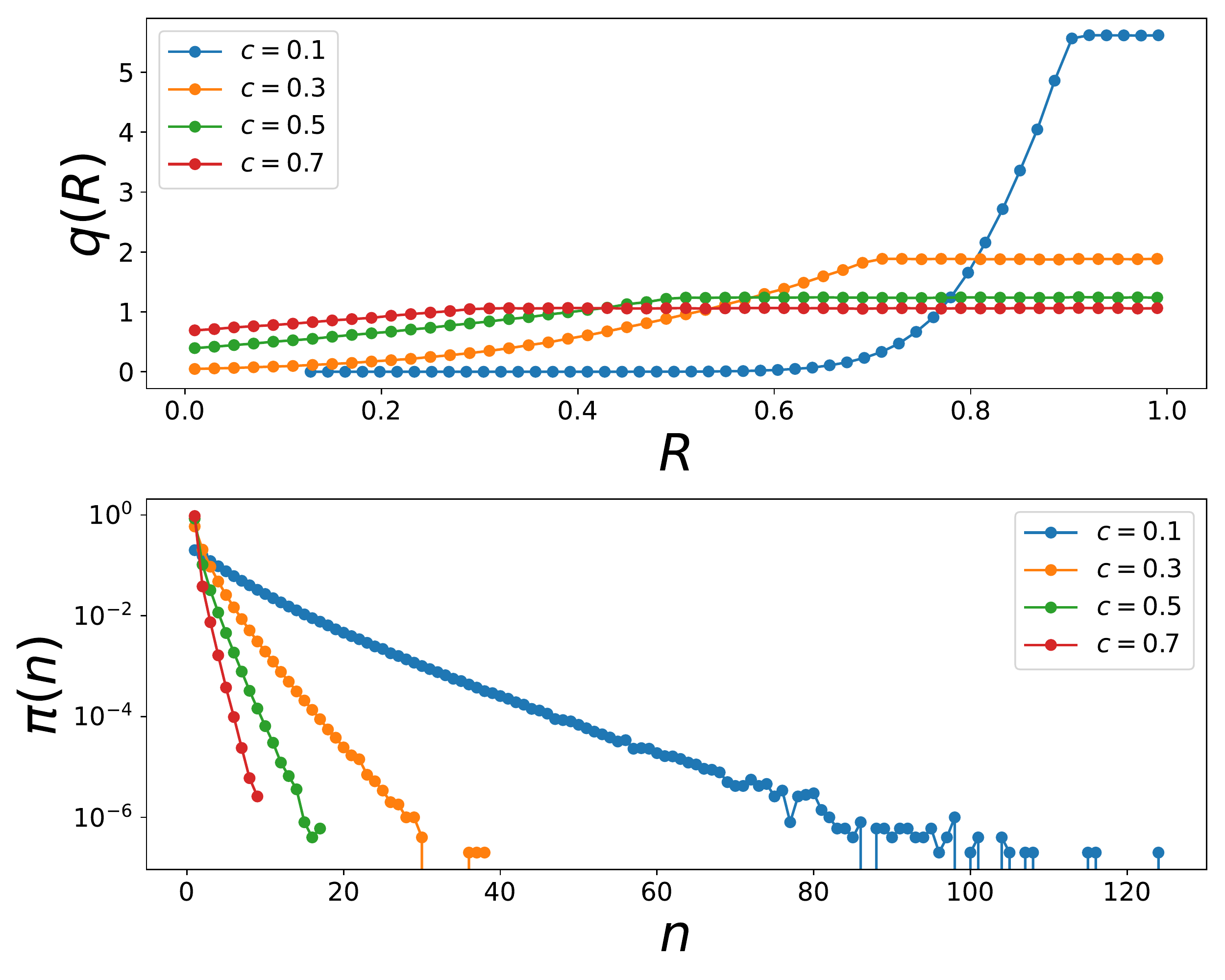}
\caption{
Record (upper panel) and age (lower panel) distributions for various $c$ 
for the uniform distribution. Analytical results 
of Eqs. (\ref{eqn:qR_unif_distr}) and (\ref{eqn:pn_unif_distr}) 
are shown by solid lines
in the upper panel.}
\label{fig:uniform_case}
\end{figure}

\subsection{Generic bounded distributions}

The $c$-record process associated to a generic bounded distribution is 
more difficult to characterize analytically. However some of the 
features that we found for the case of the uniform distribution remain 
valid. In particular for any value of $c$ the mean number of records 
grows linearly with $N$ (at large $N$) and the stationary distribution 
of ages displays an exponential cutoff. As an illustration we studied 
numerically the family of distribution wth cumulative 
$F(x)=1-(1-x)^\nu$. The uniform distribution corresponds to $\nu=1$. In 
Fig. (\ref{fig:bounded2_case}) we report our results for $\nu=2$: in the 
upper panel the study of the stationary record distribution $q(R)$ and 
in the lower panel the age distribution $\pi(n)$.

\begin{figure}[h!]
   \includegraphics[width=\linewidth]{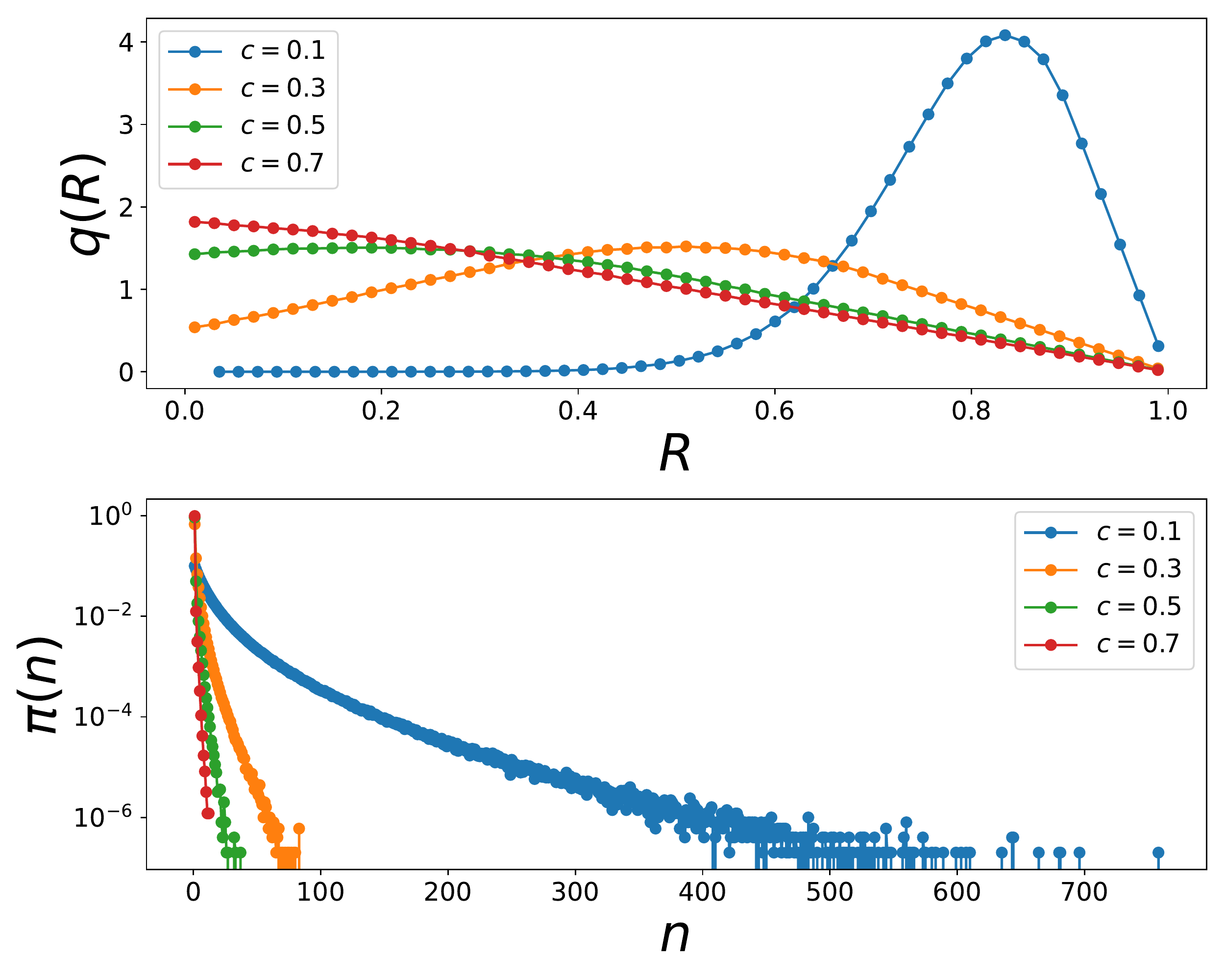}
\caption{
Record (upper panel) and ages (lower panel) distributions for various 
$c$ for the bounded case with cumulative distribution
$F(x)=1-(1-x)^{2}$ with $0\le x\le 1$.}
\label{fig:bounded2_case}
\end{figure}

\subsection{Numerical results for the Weibull family}
In this section we briefly summarize the numerical results for $\langle 
M \rangle_N$ $q(R)$, $\pi(n)$ for the Weibull family. In  
Fig. (\ref{fig:M_N_weibull}) we show the mean number of records $\langle M 
\rangle_N$ as a function of $N$ for $10^3$ realizations of the 
$c$-record process. For large enough $N$ the scaling $\langle M \rangle_N 
\propto N$ is recovered. 
In Fig.~(\ref{fig:weibull_record_distr}) we 
show the stationary record distribution for the Weibull distribution at 
$\gamma=2$. We find a stationary distribution for any $c>0$. As $c$ gets 
bigger, $q(R)$ approaches $f(R)$, as expected. In 
Fig. (\ref{fig:weibull_avg_record})
we show the average record $\langle R \rangle_\gamma (c)$ as a 
function of $c$ for different $\gamma$'s. 
We also insert the scaling as $c \to 0^+$ of the average record value 
obtained using the argument in appendix \ref{app:avg_record}:
\begin{equation}
 \label{eqn:avg_record_argument_weibull}
     \langle R \rangle_\gamma (c) \approx (c \gamma)^{\frac{1}{\gamma-1}}
 \end{equation}
Finally Fig.~(\ref{fig:gap_distr_weibull}) shows the age of record 
distributions at $\gamma=2$ for different values of $c$. The 
distributions have a power law decay as $n \to \infty$ with an exponent 
$\tau \geq 2$, i.e., $\pi(n) \sim n^{-\tau}$. This numerical result is 
compatible with the scaling $\langle M \rangle_N \propto N$ of the average 
number of records since a power law with $\tau \geq 2$ has a 
well defined first moment.
 
 \begin{figure}[h!]
   \includegraphics[width=\linewidth]{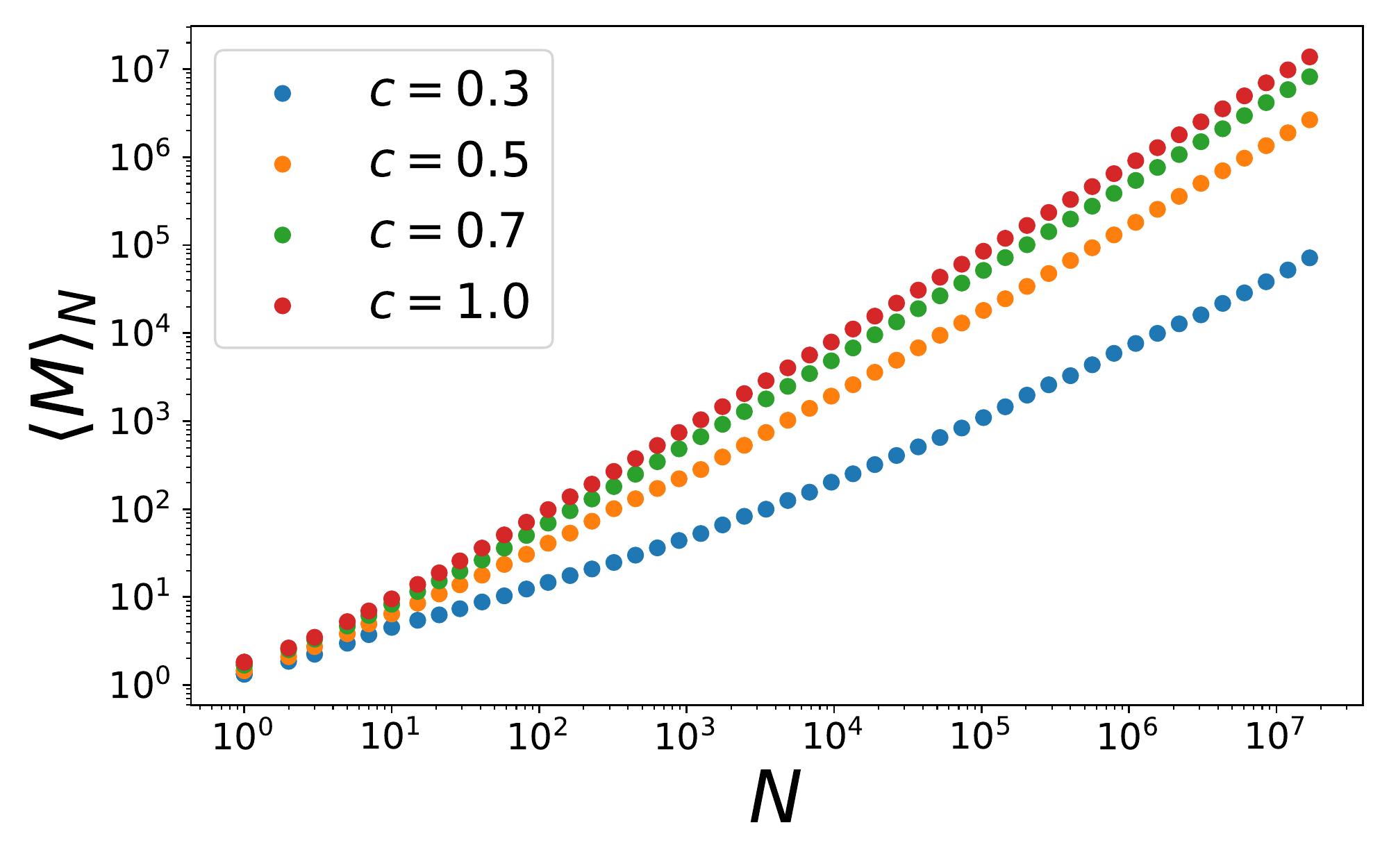}
\caption{Weibull case with $\gamma=2$: mean number of $c$-records for various $c$. As $c$ gets 
smaller, the converges to $\langle M \rangle_N \propto N$ is }
\label{fig:M_N_weibull}
\end{figure}

\begin{figure}[h!]
\includegraphics[width=\linewidth]{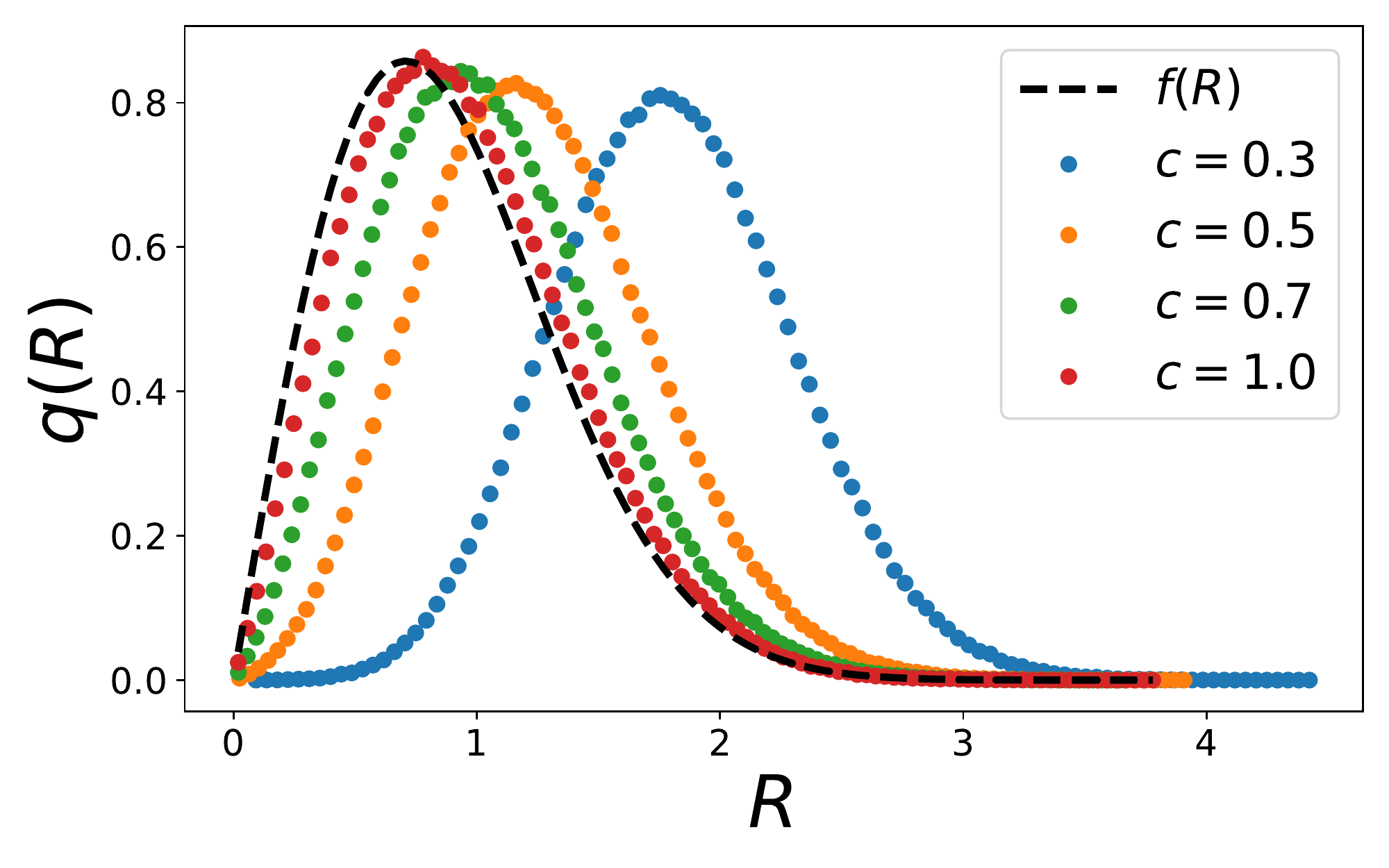}
\caption{Record distributions for the Weibull family
with $\gamma=2$ for various $c$.}
\label{fig:weibull_record_distr}
\end{figure}

\begin{figure}[h!]
   \includegraphics[width=\linewidth]{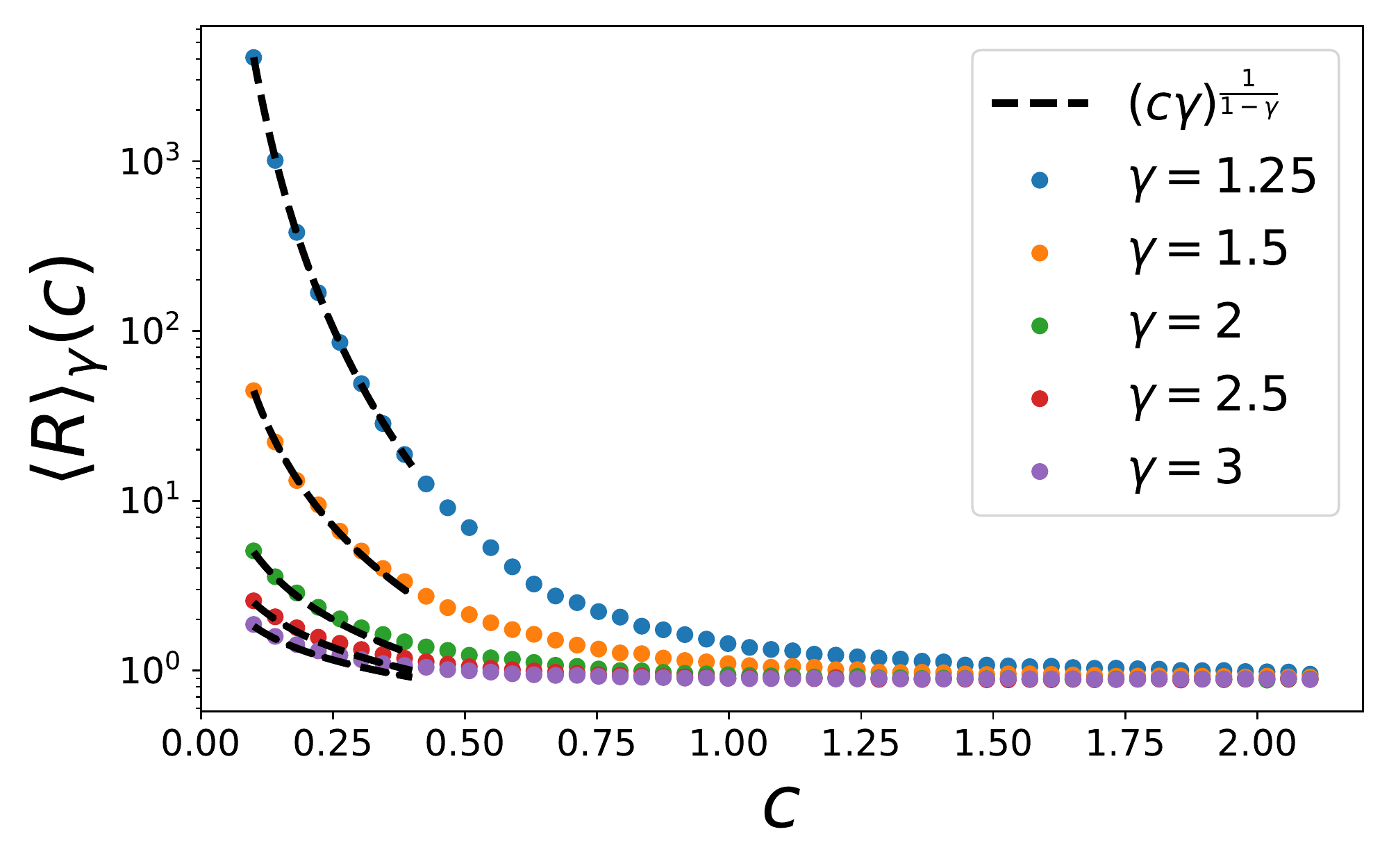}
\caption{
Average record value for various $\gamma$ and $c$ for the
Weibull family. We plot with the 
dashed black line the expected scaling of the average for $c \to 0^+$
(see Eq. (\ref{eqn:avg_record_argument_weibull})).}
\label{fig:weibull_avg_record}
\end{figure}

\begin{figure}[h!]
    \includegraphics[width=\linewidth]{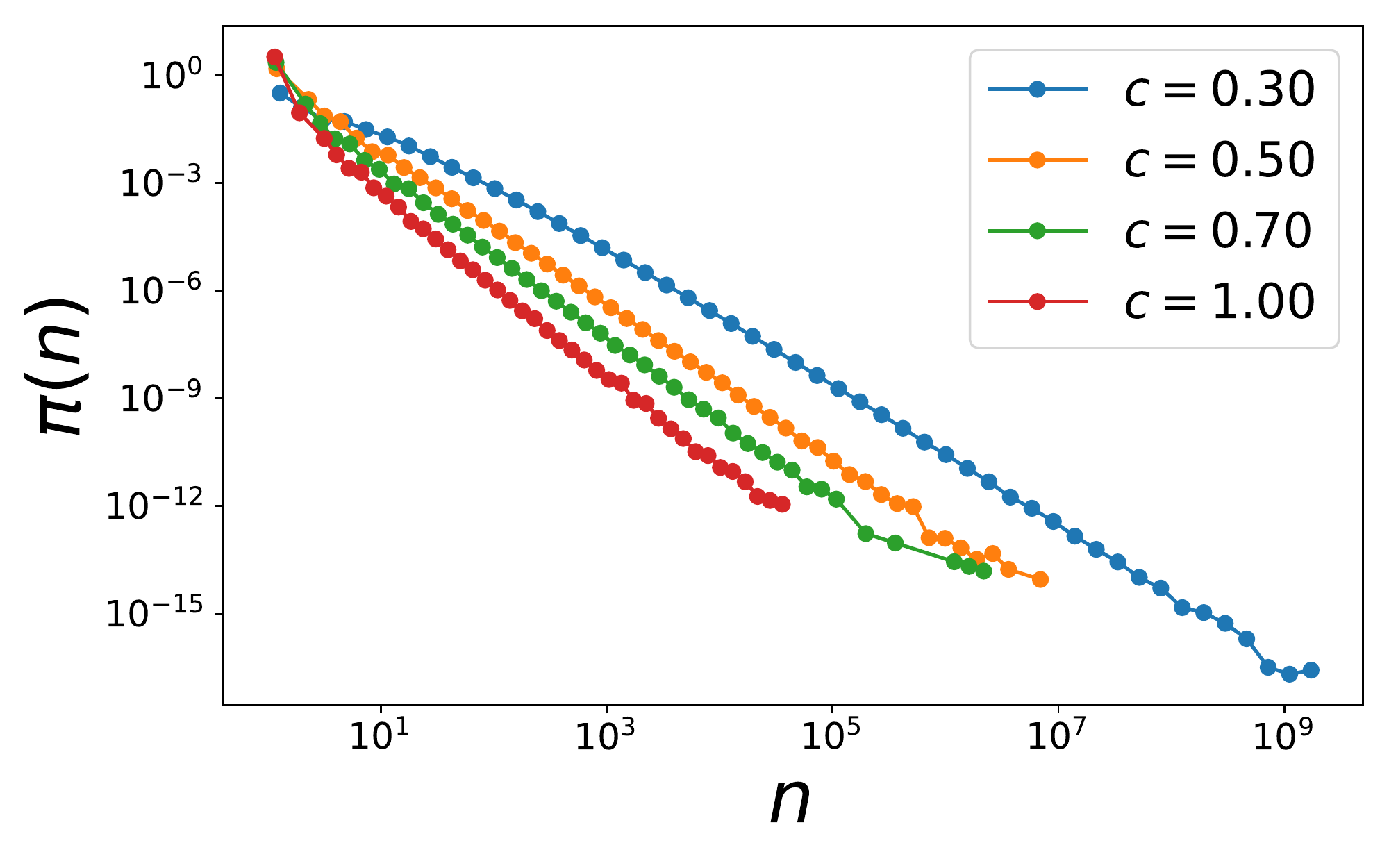}
\caption{
Age of record numerical distributions for the Weibull family at 
$\gamma=2$ and varying $c$. All the distributions show a power law tail 
with exponent $\geq 2$.}
\label{fig:gap_distr_weibull}
\end{figure}

\section{\label{sec:generalization}Generalizations of the record process}
Before the conclusion we would like to discuss some possible generalizations of the $c$-record problem. For simplicity here we focus only on the conditions for the existence of a stationary record distribution and on its form. We let the calculations of the mean number of records and of the age statistics to future works. Few protocols can be considered as a straightforward generalization of the $c$-record process:
\begin{itemize}
	\item The constant $c$ can be promoted to be a positive random variable with distribution $g(c)$. For $f(x)=e^{-x} \theta(x)$ the fixed point equation for the stationary record distribution, averaged over all possible values of  $c$ (annealed average), writes: 
\begin{equation}
    q'(R)=\int_0^\infty g(c) q(R+c) dc - q(R)
\end{equation}
Remarkably this equation admits an exponential solution
$q(R)=\lambda e^{-\lambda R}$ if the equation 
\begin{equation}
    1-\lambda = \tilde{g}(\lambda) \equiv \int_0^\infty e^{-\lambda c} g(c) dc
\end{equation}
has a positive solution $\lambda$. For example if $g(c)$ is an exponential distribution, a stationary state is reached if its mean is bigger than $1$.
\item The definition of the $c$-record can be extended with a function 
$c(R)$, namely $R_k$ is a record if $R_k > R_{k-1} - c\,(R_{k-1})$. 
As a concrete example one can consider $c(R)=c\, R$ with $c<1$ 
(for $c \geq 1$ all the values of the time series are records):
\begin{equation}
R_k > (1-c)\, R_{k-1}
\end{equation}
The fixed point equation for the stationary record distribution $q(R)$ satisfies:
\begin{flalign}
 \label{eqn:qR_sc_generalized_diff}
    & q'(R)= \frac{f(R)}{1-F(R)}  \frac{1}{1-c} q\left(\frac{R}{1-c}\right) + \\
    & +\frac{f'(R)}{f(R)} q(R) \nonumber
\end{flalign}
Equation (\ref{eqn:qR_sc_generalized_diff}) becomes simple for a Pareto distribution $f(x)=\frac{\alpha}{x^{\alpha+1}} \theta(x-1)$:
\begin{equation}
 \label{eqn:qR_sc_generalized_diff_pareto}
    q'(R)=  \frac{\alpha}{R(1-c)}  q\left( \frac{R}{1-c} \right)  - \frac{\alpha+1}{R} q(R)
\end{equation}
The stationary state exists for $c > 1-e^{-\frac{1}{\alpha}}$ and the solution of (\ref{eqn:qR_sc_generalized_diff_pareto}) still a Pareto distribution $q(R)=\frac{\beta}{R^{\beta+1}} \theta(R-1)$ with $\beta$ the unique positive solution of the trascendental equation
\begin{equation}
	1-\frac{\beta}{\alpha}=(1-c)^\beta
\end{equation}
The records generated by this process are equivalent to the $c$-records discussed in this paper via the map $R_k \to \ln R_k$.
Under this mapping the Pareto distribution becomes the exponential distribution.
\item Finally we consider the following $k$-dependent record condition:
\begin{equation}
R_k > R_{k-1} - c(k+1)^{b-1}
\end{equation} 
for a constant $b>0$. This protocol has been considered in~\cite{PK16} in 
the context of evolutionary biology: 
the quantity $c(k+1)^{b-1}$ is called \textit{handicap} and the 
analysis is carried out for both increasing $b>1$ and decreasing $b<1$ 
handicaps. The case $b=1$ coincides with the $c$-record process. 
We refer the reader to the original work \cite{PK16} for details.
\end{itemize}

\section{\label{sec:conclusion}Conclusion}

In this paper, we have shown that a simple record model of an I.I.D series, 
which we call the $c$-record model, can be successfully used to understand
and explain several realistic features of avalanche statistics 
in disordered systems, an example being the earthquake dynamics in
seismicity. This model has a single parameter $c\ge 0$ and
the other input is the distribution $f(x)$ of an entry.
We have focused on three natural observables:
(i) the mean number of records $\langle M_N\rangle$ up to step $N$
in an infinite series (ii) the distribution $q_k(R)$ of the value
of the $k$-th record and (iii) the distribution $\pi_k(n)$ of
the time interval $n$ between the $k$-th and the $(k+1)$-th record.

One of our main conclusions is that if $f(x)$ decays, for large $x$,
slower than an exponential, both 
$q_k(R)$ and $\pi_k(n)$ do not have stationary limits as $k\to \infty$
and $\langle M_N\rangle \sim \ln N$ for large $N$, as in the $c=0$ case.
Thus the effect of $c$ is not very significant for $f(x)$ with
a slower than exponential tail. In contrast, if $f(x)$ has a
faster than exponential tail, both $q_k(R)\to q(R)$ and $\pi_k(n)\to \pi(n)$ 
approach stationary limiting forms as $k\to \infty$. In particular,
we show that $\pi(n)$ decays faster than $1/n^2$ for large $n$ (indicating
that $\langle n\rangle$ is finite). Additionally, in this case,
the mean number of records grows linearly as 
$\langle M_N\rangle \sim A_1(c)\, N $ for large $N$ with $A_1(c)\le 1$. 
Thus, for $f(x)$ decaying faster than exponential, the statistics
of these three observables for finite $c$ are fundamentally different
from the standard $c=0$ case. When $f(x)$ has an exponential tail,
it turns out to be a marginal case where there is a phase transition
at a critical value $c_{\rm crit}$. For $c<c_{\rm crit}$,
the observables have qualitaively similar behavior as the $c=0$ case.
In contrast for $c>c_{\rm crit}$, both $q_k(R)$ and $\pi_k(n)$
have stationary limits as $k\to \infty$. 
We have illustrated by an explicit calculation
for $f(x)=e^{-x}\theta(x)$ for which $c_{\rm crit}=1$.
In this case we have shown that for $c>1$, the stationary avalanche
size distribution $\pi(n)\sim n^{-1-\lambda(c)}$ has a power law tail 
for large $n$ with $\lambda(c)\le 1$, indicating that the first moment diverges.
Remarkably, the exponent $\lambda(c)$ depends continuously on
$c$ and is given by the root
of the transcendental equation, $c=- \ln(1-\lambda)/\lambda$.
We have also computed exactly the stationary record value distribution $q(R)$
for $c>1$ and shown that it is a pure exponential, 
$q(R)= \lambda(c)\, e^{-\lambda(c)\, R}$, for all $R\ge 0$. 

An important feature of this $c$-record model is a nontrivial 
correlation structure between record values, as well as between 
record intevals. In this paper we have explored only partially this structure,
and it would be interesting to characterize this correlation structure
in a more complete fashion.

We have also provided some generalisations of this simple $c$-record model,
where the criteria for record formation, i.e., the offset $c_k$ 
depends on the record value $R_k$ as well as on the record index $k$.
In all these cases, the offset $c_k$ remains constant (albeit $k$-depenent)
as in the lower panel of Fig. (\ref{fig:schema}). Previously,
the linear trend model was studied where the offset decreases linearly
with time during an avalanche (as in the upper panel of 
Fig. (\ref{fig:schema})). One can then ask how the record statistics
gets modified for a general decreasing offset function during an
avalanche. 

Finally, in this paper, we have considered a series of I.I.D variables 
as a model for the pinning force landscape. It is natural to
consider a model where the landscape is correlated. For example, in
the ABBM model, the entries of the series correspond to the positions of a
random walker. This remains a challenging open problem.

\acknowledgements

We thank O. Giraud for an illuminating discussion. We are grateful to J. Krug
and S.-C. Park for very useful exchanges and correspondences.

\pagebreak

\appendix
\section{\label{app:exp_case_PNM}Exponential case: $P_N(M,R)$}
Our starting point is Eq. (\ref{eqn:QR_exp_case})
for the generating function ${\tilde Q}(R)$ in
the exponential case, $f(x)=e^{-x}\, \theta(x)$.
We solve it in terms of $H(R)=\int_0^R \tilde{Q}(R') dR'$. 
Thus Eq. (\ref{eqn:QR_exp_case}) becomes:
\begin{multline}
\label{eqn:HR_diff_equation}
    H'(R) \left[1-z(1-e^{-(R-c)}) \theta(R-c) \right]  = \\
    z\,s\, e^{-R}\, \left[1+ H(R+c)\right] 
\end{multline}
The initial condition for $H(R)$ is $H(0)=0$. In order to solve 
(\ref{eqn:HR_diff_equation}) we need to distinguish region I with $R<c$, 
region II with $R>c$ and match the two solutions at $R=c$. We call 
$H_I(R)$ the solution for region I and $H_{II}(R)$ the one for region 
II. 

We start with region II ($R>c$) and use the ansatz
\begin{equation}
\label{eqn:HII_diffeq}
    H_{II}(R)=\sum_{m=0}^\infty a_m e^{-m\,R}
\end{equation}
Substituting this in Eq. (\ref{eqn:HR_diff_equation}) and
matching terms of $e^{-R}$ on both sides, we get 
the following recursion relation for the coefficients $a_m$'s
\begin{flalign}
    & a_1 = - \frac{zs}{1-z}(1+a_0) \\
    & a_m = -\frac{ze^c}{1-z} \left(\frac{m-1+se^{-mc}}{m}\right) a_{m-1} \, .
\end{flalign}
By iterating we get
\begin{equation}
    a_m=(-1)^m \left(\frac{ze^c}{1-z}\right)^m \frac{1}{m!} b_m(s) (1+a_0)
\end{equation}
where
\begin{equation}
\label{eqn:bms}
    b_m(s) = \begin{cases}
    \prod_{k=1}^m (k-1+se^{-kc}) & m \geq 1 \\
    1 & m=0
    \end{cases}
\end{equation}
We then rewrite $H_{II}(R)$ as
\begin{equation}
    H_{II}(R) = (1+a_0) \sum_{m=0}^\infty 
\left(-\frac{ze^c}{1-z}\right)^m \frac{b_m(s)}{m!} e^{-mR}\, .
\label{HII.1}
\end{equation}
The coefficient $a_0$ is a parameter that must be set by 
imposing the matching of $H_I(R)$ and $H_{II}(R)$ at $R=c$. 
In the region I, $H_I(R)$ satisfies:
\begin{equation}
\label{eqn:HI_diffeq}
    (1-z) \frac{d H_I(R)}{dR} = zs e^{-R} (1+H_{II}(R+c))
\end{equation}
Substituting the solution of $H_{II}(R)$ from Eq. (\ref{HII.1})
on the r.h.s of (\ref{eqn:HI_diffeq}) and solving
the first order equation for $H_I(R)$, using $H_I(0)=0$, gives
\begin{multline}
    H_I(R) = sz (1+a_0) \sum_{m=0}^\infty \left(-\frac{z}{1-z}\right)^m \\
    \frac{b_m(s)}{(m+1)!} \frac{(1-e^{-(m+1)R})}{m+1}
\end{multline}
The continuity at $R=c$, namely $H_I(c)=H_{II}(c)$, fixes $a_0$:
\begin{equation}
\label{eqn:a0_exp_first_form}
    a_0 = \frac{1}{\sum_{m=0}^\infty \left(-\frac{z}{1-z}\right)^m \frac{b_m(s)}{m!} \left(1-\frac{zs(1-e^{-(m+1)c})}{m+1}\right)}-1
\end{equation}

In the limit $R\to \infty$, we have 
$H(\infty)=\int_0^{\infty} {\tilde Q}(R')\, dR'$.
On the other hand, it follows by taking $R\to \infty$ limit
in Eq. (\ref{eqn:HII_diffeq}) that $H(\infty)=a_0$.
Combining these two and using
Eq.~(\ref{eqn:PNM_definition_gf}), we finally identify:
\begin{equation}
    \sum_{N=1}^\infty \sum_{M=1}^\infty P_N(M) z^N s^M = a_0 = a_0(z,s)\, ,
\end{equation}
with $a_0$ given in Eq. (\ref{eqn:a0_exp_first_form}).
One can check that for $c=0$
\begin{equation}
    a_0=\frac{1}{(1-z)^s}-1
\end{equation}
which reproduces the well known result for the generating function of the 
standard record process~\cite{GMS17}. 
To find the generating function of $\langle M \rangle_N$ one must 
study $a_0$ in the vicinity of $s=1$. 
The expansion around $s=1$ from (\ref{eqn:a0_exp_first_form}) is not straightforward. We want to rewrite the denominator of $a_0$ in a simpler form using $\tilde{b}(x,s)$, the generating function of $b_m(s)$. We first introduce an useful identity:
\begin{flalign}
\label{eqn:bstilde_0}
    & \tilde{b}(x,s)=\sum_{m=0}^\infty \frac{b_m(s)}{m!} x^m \nonumber \\
    & =1 + \sum_{m=1}^\infty (m-1+se^{-mc}) \frac{b_{m-1}(s)}{m!} x^m = \nonumber \\ 
    & = 1 + \sum_{m=1}^\infty \frac{b_{m-1}(s) x^m}{(m-1)!}-\sum_{m=1}^\infty \frac{b_{m-1}x^m}{m!} + \nonumber \\
    & + s \sum_{m=1}^\infty \frac{b_{m-1}(xe^{-c})^m}{m!} \nonumber \\
    & = 1 + x \tilde{b}(x,s)-\sum_{m=1}^\infty \frac{b_{m-1}x^m}{m!} + \nonumber \\
    & + s \sum_{m=1}^\infty \frac{b_{m-1}(xe^{-c})^m}{m!}\, .
\end{flalign}
Eq. (\ref{eqn:bstilde_0}) can be rewritten as:
\begin{equation}
\label{eqn:bstilde_1}
    (1-x) \tilde{b}(x,s) = 1-\sum_{m=1}^\infty \frac{b_{m-1}(s)x^m}{m!} + s \sum_{m=1}^\infty \frac{b_{m-1}(s)(xe^{-c})^m}{m!}
\end{equation}
We now identify $x=-z/(1-z)$ and thus $z=-x/(1-x)$. Using Eq. 
(\ref{eqn:bstilde_1}), the denominator of Eq. (\ref{eqn:a0_exp_first_form}) 
can be rewritten as:
\begin{flalign}
    & \sum_{m=0}^\infty \frac{x^m}{m!} b_m(s) + 
\frac{xs}{1-x} \sum_{m=0}^\infty \frac{b_m(s) x^m}{m! (m+1)} 
\left[1-e^{-(m+1)c}\right] \nonumber \\
    & = \tilde{b}(x,s) + \frac{xs}{1-x} \sum_{m=0}^\infty 
\frac{b_m(s) x^m}{(m+1)!}-\frac{xse^{-c}}{1-x} 
\sum_{m=0}^\infty \frac{b_m(s) (xe^{-c})^m}{(m+1)!} \nonumber \\
    & = \tilde{b}(x,s) + \frac{s}{1-x} \sum_{m=1}^\infty \frac{b_{m-1}x^m}{m!} + \nonumber \\
    & - \frac{1}{1-x} \left[(1-x) \tilde{b}(x,s)-1+\sum_{m=1}^\infty \sum_{m=1}^\infty \frac{b_{m-1} x^m}{m!}\right] \nonumber \\
    & = \frac{1}{1-x}\left[1-(1-s) \sum_{m=1}^\infty 
\frac{b_{m-1}(s)x^m}{m!} \right] 
\end{flalign}
Reintroducing $z=-x/(1-x)$, we rewrite Eq.~(\ref{eqn:a0_exp_first_form}) in a 
more amenable form to carry out the expansion around $s=1$:
\begin{equation}
\label{eqn:a0_exp_second_form}
   1+ a_0 = \frac{1}{(1-z) \left[1-(1-s)\sum_{m=1}^\infty \frac{b_{m-1}(s)}{m!} \left(-\frac{z}{1-z}\right)^m\right]}
\end{equation}
By expanding $a_0$ in $s=1-\epsilon$ as $\epsilon \to 0$ we can 
extract the derivative of $a_0$ in $s=1$. After some algebra we finally arrive at the generating function of $\langle M \rangle_N$:
\begin{multline}
\label{eqn:genfunc_MN}
    \tilde{M}(z) = \sum_{N=1}^\infty \langle M \rangle_N z^N = \left. \frac{\partial a_0}{\partial s} \right|_{s=1} = \\
    = -\frac{1}{1-z} \sum_{m=0}^\infty \frac{b_m(1)}{(m+1)!} \left(-\frac{z}{1-z}\right)^{m+1}
\end{multline}
In accordance to Eq.~(\ref{eqn:MN_by_derivatives}), the average number of 
records $\langle M \rangle_N$ identifies with the coefficient $z^N$ in the 
series expansion of $\tilde{M}(z)$ at $z=0$. The explicit expression of 
$\langle M \rangle_N$ is given in Eq.~(\ref{eqn:exp_MN}).

\section{\label{app:exp_case_MN_asymp}Asymptotics of $\langle M \rangle_N$}
From now on $b_m = b_m(1)$ and $\tilde{b}(x)=\tilde{b}(x,1)$. 
The behavior of $ \langle M \rangle_N$ for $N \to \infty$ 
is hard to extract from Eq.~(\ref{eqn:exp_MN}). It is then useful to 
restart from Eq.~(\ref{eqn:genfunc_MN}) and introduce the 
function $\tilde{g}(x)$ such that:
\begin{equation}
\label{eqn:genfunc_MN_gtilde}
    \tilde{M}(z) = - \frac{1}{1-z} \tilde{g} \left(-\frac{z}{1-z}\right)
\end{equation}
The function $\tilde{g}(x)$ writes:
\begin{equation}
\label{eqn:gtilde_definition}
    \tilde{g}(x) = \sum_{m=0}^\infty b_m \frac{x^{m+1}}{(m+1)!} 
\end{equation}
The asymptotics of $\langle M \rangle_N$ for large $N$ is 
controlled by the behavior of $\tilde{M}(z)$ when $z \to 1$, 
which is equivalent to $\tilde{g}(x)$ when $x \to -\infty$.  
By the definition of $\tilde{b}(x)$ (\ref{eqn:bstilde_0}) 
it is straightforward to note that:
\begin{equation}
\label{eqn:gtilde_derivative}
    \frac{d \tilde{g}(x)}{dx} = \tilde{b}(x)
\end{equation}
Setting $s=1$ in Eq.~(\ref{eqn:bstilde_1}) gives
\begin{equation}
\label{eqn:bstilde_1_at_s=1}
    (1-x) \tilde{b}(x) = 1-\sum_{m=1}^\infty \frac{b_{m-1}x^m}{m!} 
+ \sum_{m=1}^\infty \frac{b_{m-1}(xe^{-c})^m}{m!}\, .
\end{equation}
We can derive two differential equation for $\tilde{g}(x)$ and $\tilde{b}(x)$:
The first is obtained by rewriting each term of (\ref{eqn:bstilde_1_at_s=1}) using $\tilde{g}(x)$:
\begin{equation}
\label{eqn:gtilde_diffeq}
    (1-x)\frac{d \tilde{g}(x)}{dx} = 1-\tilde{g}(x) + \tilde{g}(xe^{-c})
\end{equation}
The second is obtained from the derivative of (\ref{eqn:bstilde_1_at_s=1}) with respect to $x$: 
\begin{equation}
\label{eqn:btilde_diffeq}
    \frac{d \tilde{b}(x)}{dx} = \frac{e^{-c} \tilde{b}(e^{-c}x)}{1-x}
\end{equation}
Note that Eq.~(\ref{eqn:btilde_diffeq}) is a first order  
non-local differential equation. The representation of 
$\tilde{b}(x)$ given in Eq.~(\ref{eqn:bstilde_0}) is not suitable 
for studying the limit $x \to -\infty$. Instead the solution of 
(\ref{eqn:btilde_diffeq}) gives us a nice series representation of 
$\tilde{b}(x)$ around $x \to -\infty$ using the following ansatz: 
\begin{equation}
\label{eqn:btilde_ansatz}
    \tilde{b}(x) = \sum_{k=0}^\infty A_k (-x)^{-\phi - k}
\end{equation}
By plugging the ansatz in (\ref{eqn:btilde_diffeq}) we find that coefficients satisfy:
\begin{equation}
\label{eqn:Ak_recurrence_coefs}
    A_k = \frac{k-1+\phi}{e^{ck}\phi - k - \phi} A_{k-1}
\end{equation}
with $A_0$ to be determined. The exponent $\phi$ is the solution of:
\begin{equation}
\label{eqn:phi_selfcons}
    \phi = (e^{-c})^{1-\phi}
\end{equation}
At most two solutions exist for (\ref{eqn:phi_selfcons}).
\begin{itemize}
    \item For $c<1$ one solution is $\phi=1$ and the other is $\phi(c)>1$.
    \item For $c>1$ one solution is $\phi=1$ and the other is $\phi(c)<1$
    \item For $c=1$ the two solutions coincide i.e. $\phi=1$.
\end{itemize}
From now on we assume $c \neq 1$ and rewrite the ansatz (\ref{eqn:btilde_ansatz}) in order to make explicit the existence of two solutions:
\begin{equation}
\label{eqn:btilde_ansatz_solution}
    \tilde{b}(x) = A_0 \sum_{k=0}^\infty \frac{c_k}{(-x)^{\phi+k}} + B_0 \sum_{k=0}^\infty \frac{\tilde{c}_k}{(-x)^{k+1}}
\end{equation}
where:
\begin{equation}
    c_k = \prod_{m=1}^k \frac{m-1+\phi}{(e^{cm}-1)\phi - m}
\end{equation}
and:
\begin{equation}
    \tilde{c}_k = \prod_{m=1}^k \frac{m}{e^{cm}- m-1}
\end{equation}
with $c_0=\tilde{c}_0=1$. Both $c_k$ and $\tilde{c}_k$ come from the iteration of equation (\ref{eqn:Ak_recurrence_coefs}).
To complete the solution we have to find $A_0$ and $B_0$. 
Before that we note that identifying $\phi(c)=1-\lambda(c)$ equation 
(\ref{eqn:exp_c_lmbd}) is equivalent to (\ref{eqn:phi_selfcons}). The 
case $c>1$ corresponds to a positive $\lambda(c)$, to the existence of a 
stationary state and when $x \to -\infty$ to $\tilde{b}(x) \sim A_0 
(-x)^{-\phi}$. While when $c < 1$ no stationary state exists and 
$\tilde{b}(x) \sim B_0/(-x)$. \\

To set $A_0$ and $B_0$ we need two relations. The first relation is found by matching at $x=-1$ the value of $\tilde{b}(x)$ in the two series representation given by (\ref{eqn:bstilde_0}) and (\ref{eqn:btilde_ansatz_solution}): 
\begin{equation}
\label{eqn:A0B0_solution}
    \tilde{b}(-1) = \sum_{m=0}^\infty (-1)^m \frac{b_m}{m!} = A_0 \sum_{k=0}^\infty c_k + B_0 \sum_{k=0}^\infty \tilde{c}_k
\end{equation}
From the solution (\ref{eqn:btilde_ansatz_solution}) we can find, up to an integration constant $\mu(c)$, $\tilde{g}(x)$:
\begin{flalign}
\label{eqn:btilde_integrated_to_gtilde}
    & \tilde{g}(x) = A_0 \sum_{k=0}^\infty 
\frac{c_k(-x)^{-\phi -k +1}}{\phi + k - 1} + B_0 (-\ln(-x)) + \nonumber \\
   & + B_0 \sum_{k=1}^\infty \tilde{c}_k  \frac{(-x)^{-k}}{k} + \mu(c) \, . 
\end{flalign}
The second relation corresponds to imposing that 
Eq.~(\ref{eqn:btilde_integrated_to_gtilde}) must be a solution of 
(\ref{eqn:gtilde_diffeq}):
\begin{multline}
    (1-x) \left[A_0 \sum_{k=0}^\infty c_k(-x)^{-\phi -k}+B_0 \sum_{k=0}^\infty \tilde{c}_k(-x)^{-k-1}\right] = \\
    = 1- \left[A_0 \sum_{k=0}^\infty \frac{c_k(-x)^{-\phi -k +1}}{\phi + k - 1} + B_0 (-\ln(-x)) \right]+ \\
     - \left[ B_0 \sum_{k=1}^\infty \tilde{c}_k  \frac{(-x)^{-k}}{k} + \mu(c)\right] + \\
    + \left[A_0 \sum_{k=0}^\infty \frac{c_k(-xe^{-c})^{-\phi -k +1}}{\phi + k - 1} + B_0 (-\ln(-x e^{-c})) \right] + \\
     + \left[B_0 \sum_{k=1}^\infty \tilde{c}_k  \frac{(-x e^{-c})^{-k}}{k} + \mu(c)\right]
\end{multline}
$B_0$ comes from matching the coefficients of order $(-x)^0$:
\begin{equation}
    B_0 = \frac{1}{1-c}
\end{equation}
By matching coefficients $(-x)^{-k}$ and $(-x)^{-\phi - k}$ one confirms the solution (\ref{eqn:btilde_ansatz_solution}). $A_0$ is determined 
using Eq.~(\ref{eqn:A0B0_solution}):
\begin{equation}
\label{eqn:A0_solution}
    A_0 = \frac{\sum_{m=0}^\infty (-1)^m \frac{b_m}{m!}-\frac{1}{1-c}\sum_{k=0}^\infty \tilde{c}_k}{\sum_{k=0}^\infty c_k}
\end{equation}
Finally we can write $\tilde{g}(x)$:
\begin{multline}
    \tilde{g}(x) = A_0 \sum_{k=0}^\infty \frac{c_k(-x)^{-\phi -k +1}}{\phi + k - 1} + \frac{1}{1-c} \left[-\ln(-x)\right] + \\
    + \frac{1}{1-c} \sum_{k=1}^\infty \tilde{c}_k  \frac{(-x)^{-k}}{k} + \mu(c)
\end{multline}
The constant $\mu(c)$ is up to now undetermined and must be found by matching $\tilde{g}(x)$ at $x=-1$ with the power series representation in (\ref{eqn:gtilde_definition})
\begin{flalign}
    & A_0 \sum_{k=0}^\infty \frac{c_k}{\phi + k - 1} + \frac{1}{1-c} \sum_{k=1}^\infty  \frac{\tilde{c}_k}{k} + \mu(c)  = \\
    & = \sum_{m=0}^\infty \frac{b_{m}(-1)^{m+1}}{(m+1)!} \nonumber
\end{flalign}
Having now a representation for $\tilde{g}(x)$ we can invert 
$\tilde{M}(z)$ by making use of Eq.~(\ref{eqn:genfunc_MN_gtilde}).  For $z \to 1$ it is useful approximate the series of the generating function with a Laplace transform by setting $z = 1-s$ in the limit $s \to 0$:
\begin{equation}
    \sum_{N=1}^\infty \langle M \rangle_N z^N \approx \int dN \langle M \rangle_N e^{-sN} 
\end{equation}
Using equation (\ref{eqn:genfunc_MN_gtilde}):
\begin{equation}
    \mathcal{L}_s \left[\langle M \rangle_N \right] =  \int dN \langle M \rangle_N e^{-sN}  \approx - \frac{1}{s} \tilde{g} \left(-\frac{1}{s}\right)
\end{equation}
We now make use of two well known inverse Laplace transforms:
\begin{flalign}
\label{eqn:invlapl_MN}
    & \mathcal{L}_s^{-1} \left[\frac{1}{s^\alpha}\right] = \frac{1}{\Gamma(\alpha)} N^{\alpha-1} \\
    & \mathcal{L}_s^{-1} \left[-\frac{\ln(s)}{s}\right] = \ln(N)
\end{flalign}
For $c>1$ we find:
\begin{equation}
    \langle M \rangle_N \approx \frac{A_0}{(1-\phi)\Gamma(2-\phi)} N^{1-\phi} + \frac{1}{1-c} \ln N + O(\ln N)
\end{equation}
For $c<1$ we find:
\begin{equation}
    \langle M \rangle_N \approx \frac{1}{1-c} \ln N - \mu(c)
\end{equation}
The particular values for $A_0(c)$ and $\mu(c)$ used in the main text 
were found with Mathematica. \\

What is left now is the $c=1$ case. Since the two solutions 
for $\phi$ from (\ref{eqn:phi_selfcons}) coincide at $c=1$, 
we need to solve for $\tilde{b}(x)$ using a different `degenerate' ansatz. 
In this case Eq.~(\ref{eqn:btilde_diffeq}) becomes:
\begin{equation}
    \frac{d\tilde{b}(x)}{dx} = \frac{1}{1-x} \frac{1}{e} \tilde{b}\left(\frac{x}{e}\right)
\end{equation}
We use the following ansatz:
\begin{equation}
\label{eqn:btilde_ansatz_c_1}
    \tilde{b}(x) = \ln(-x) \sum_{k=1}^\infty A_k (-x)^{-k} + \sum_{k=1}^\infty B_k(-x)^{-k}
\end{equation}
The coefficients $A_k$ satify:
\begin{equation}
    A_k = \frac{k-1}{e^{k-1}-k} A_{k-1}
\end{equation}
with $A_1$ undetermined. The expression of the coefficients $B_k$ is more involved:
\begin{equation}
    A_k(e^{k-1}-1) - B_k(e^{k-1}-1) = A_{k-1} -(k-1) B_{k-1}
\end{equation}
with $B_1$ undetermined. The function $\tilde{g}(x)$ is obtained integrating (\ref{eqn:btilde_ansatz_c_1}). We report only the first terms:
\begin{equation}
   \tilde{g}(x) = \mu(c) -\frac{A_1}{2} \ln^2(-x) + \dots
\end{equation}
By using equations (\ref{eqn:bstilde_1_at_s=1}) and (\ref{eqn:gtilde_diffeq}) 
we can find the undetermined coefficients as we did for the $c \neq 1$ case. 
By inverting the Laplace transform we thus finally obtain the leading
asymptotic behavior for large $N$:
\begin{equation}
    \langle M \rangle_N \approx \ln^2(N)\, \quad c=1\, . 
\end{equation}

\section{\label{app:corr_exp} Exponential case: Record correlations}
By using Eqs.~(\ref{eqn:R_cond_prop}) and (\ref{eqn:exp_qR}) we can find 
the correlation between two subsequent records at the stationary state. 
We denote by $R'$ the record and its successive by $R$. First of all we compute the conditional average of $R$ given $R'$. 
It is straightforward since we we know $q(R|R')$ from equation (\ref{eqn:R_cond_prop}):
\begin{equation}
    \mathbb{E}[R|R'] = 1+ (R'-c) \theta(R'-c)
\end{equation}
The expected value of $R R'$ is thus given by:
\begin{equation}
    \mathbb{E}[R R'] = \int dR' q(R') R' \mathbb{E}[R|R']
\end{equation}
By substituting $q(R')=\lambda e^{-\lambda R'}$:
\begin{equation}
    \mathbb{E}[R R'] = \frac{1}{\lambda} + \frac{(1-\lambda)}{\lambda^2} \left(2-\ln(1-\lambda)\right)
\end{equation}
By subtracting $\mathbb{E}[R] \mathbb{E}[R'] = \lambda^{-2}$ we obtain the covariance between $R$ and $R'$:
\begin{equation}
    \text{Cov}[RR'] = \frac{1-\lambda}{\lambda^2} (1-\ln(1-\lambda))
\end{equation}
Finally the Pearson correlation coefficient takes a simple form:
\begin{equation}
    \rho(\lambda) = \frac{\text{Cov}[RR']}{\text{Var}[R']} = (1-\lambda) (1-
\ln(1-\lambda))
\end{equation}
As $\lambda \to 1$ (or $c \to 1$) $\rho(\lambda)$ diverges and we recover the long range correlation of the classical record process. On the other hand if $\lambda \to 1$ (i.e. $c \to \infty$) $\rho(\lambda) \to 0$ and all records become uncorrelated.

\section{\label{app:bounded_case_PNM}Bounded case: $P_N(M,R)$}
We restrict to the case of bounded distribution in $[0,1]$ and $c \geq 1/2$. For $c \leq R \leq 1$:
\begin{equation}
    \tilde{Q}(R) = \frac{zs(1+H)\nu (1-R)^{\nu-1}}{1-z(1-(1+c-R)^\nu)}
\end{equation}
where we denote by $H=\int_0^1 \tilde{Q}(R') dR'$. For $1-c \leq R \leq c$:
\begin{equation}
    \tilde{Q}(R) = zs(1+H) \nu (1-R)^{\nu-1}
\end{equation}
For $0 \leq R \leq 1-c$:
\begin{multline}
    \tilde{Q}(R) = zs \left( 1+ \int_0^{R+c} dR' \tilde{Q}(R') \right) =\\
    = zs(1+K) + zs \int_c^{R+c} dR' \tilde{Q}(R')
\end{multline}
where $K=\int_0^c \tilde{Q}(R') dR'$. Finally:
\begin{multline}
    \tilde{Q}(R) = zs(1+K) +  \\ 
    + zs\int_c^{R+c}  \frac{zs(1+H)\nu (1-R')^{\nu-1}}{1-z(1-(1+c-R')^\nu)} dR'
\end{multline}
$\tilde{Q}(R)$ has a simple expression in the case of uniform distribution i.e. $\nu=1$. In this case:
\begin{equation}
    \tilde{Q}(R)=zs(1+K) -zs^2 (1+H) \ln \left(1-zR\right)
\end{equation}
Now we need to impose self-consistently that
\begin{equation}
    H = \int_0^1 \tilde{Q}(R) dR
\end{equation}
We get:
\begin{equation}
    \frac{H}{1+H} = s(s-1) \ln(1-(1-c)z) + sz((1-c)s+s)
\end{equation}
The generating function of $P_N(M)$ is precisely $H$. The generating function of $\langle M \rangle_N$, namely:
\begin{equation}
    \tilde{M}(z) = \sum_{N=1} \langle M \rangle_N z^N
\end{equation}
is obtained by deriving $H$ by $s$ in $s=1$. It is easy to show that:
\begin{equation}
    \tilde{M}(z)=\frac{2-c+\ln(c)}{(1-z)^2} + \frac{z(2-c)}{(1-z)^2}
\end{equation}
By expanding around $z=0$ we recover equation (\ref{eqn:exp_MN_uniform}).

\section{\label{app:classical_record}Classical records distribution}
In the case of classical record statistics of i.i.d. continuous random variables it is straightforward to compute the distribution of the $k$-th record value in the limit of an infinite sequence. Equation (\ref{eqn:recurrence_qRk}) at $c=0$ becomes a local equation with a simple solution:
\begin{equation}
\label{eqn:qRk_c_zero}
    q_k(R)= \frac{\left[-\ln(1-F(R))\right]^{k-1}}{(k-1)!} f(R)
\end{equation}
As a reference, equation (\ref{eqn:qRk_c_zero}) has been used in \cite{RP06} for modelling record temperatures in the cases of exponential and Gaussian distributions. \\
As an explicit example, in the case of a Weibull distribution, $q_k(R)$ has a simple expression:
\begin{equation}
\label{eqn:qRk_c_zero_weibull}
q_k(R)=\frac{\gamma}{(k-1)!} R^{\gamma k - 1} e^{-R^\gamma}
\end{equation}
The average record $\langle R_k \rangle_\gamma$ can be exactly computed:
\begin{equation}
    \langle R_k \rangle_\gamma = \frac{\Gamma(k+\frac{1}{\gamma})}{(k-1)!} \to k^{1/ \gamma}\quad k \to \infty
\end{equation}
As expected, $\langle R_k \rangle_\gamma$ grows unboundendly as $k \to \infty$.

\section{\label{app:avg_record}Average record}
In this section we lay down a qualitative argument for estimating the average record $\bar{x}$ as $c \to 0^+$. The equation satisfied by $q(x)$ is:
\begin{equation}
	q(x)=f(x) \int_0^c q(y) dy + f(x) \int_c^{x+c} \frac{q(y)}{1-F(y-c)} dy
\end{equation}
We want to extract the average value. Multiply both sides by $x$ and integrate by $x$ from $0$ to $\infty$:
\begin{equation}
	\label{eqn:avg_record_recurrence}
	\bar{x} = \int_0^\infty x f(x) \int_0^c q(y) dy +
	\int_0^\infty x f(x) \int_c^{x+c} \frac{q(y)}{1-F(y-c)} dy
\end{equation}
We are interested in the limit $c \to 0^+$ and we know that in this limit the typical records become large are narrow distributed. We then try to solve equation (\ref{eqn:avg_record_recurrence}) using the ansatz $q(x)=\delta(x-\bar{x})$  with $\bar{x} \gg c \to 0^+$. 
\begin{equation}
	\bar{x} = \frac{1}{1-F(\bar{x}-c)} \int_{\bar{x}-c}^\infty xf(x) dx
\end{equation}
Here we used that $\int_0^c q(y) dy = 0$ since $\bar{x}>c$. We now take a derivative with respect to $\bar{x}$ and rearranging the terms:
\begin{equation}
	\label{eqn:hazard_1_over_c}
	\frac{f(\bar{x}-c)}{1-F(\bar{x}-c)}=\frac{1}{c}
\end{equation}
Equation (\ref{eqn:hazard_1_over_c}) is an implicit equation for $\bar{x}$ in the limit $c \to 0^+$.

If the ratio $f(x)/(1-F(x))$ is, for large $x$, an increasing function of $x$, we expect a diverging $\bar{x}$ as $c \to 0^+$, which is compatible with the classical record problem. Otherwise if $\bar{x}$ vanishes as $c \to 0^+$ we do not recover the classical record case behaviour hence we expect no stationary record distribution in this case.

\section{\label{app:simulation}Record and age of record simulation}
In order to generate $c$-records and their ages we can make use of equation (\ref{eqn:R_cond_prop}). We start from the first record, $R_1$, extracted from $f(R)$. The subsequent records have to satisfy:
\begin{equation}
    F(R_k)=u_k+(1-u_k) F(R_{k-1}-c)
\end{equation}
where $u_k$ is a random number in $[0,1]$. For any random variable for which the inversion method is feasible one can use:
\begin{equation}
    R_k=F^{-1}\left(u_k+(1-u_k) F(R_{k-1}-c)\right)
\end{equation}
Indeed in the case of Weibull distirbutions:
\begin{flalign}
    & F(x) = 1-e^{-x^\gamma} \\
    & F^{-1}(u) = (-\ln(1-u))^{1/\gamma}
\end{flalign}
which implies:
\begin{equation}
    R_k=\left[-\ln(1-u_k) + (R_{k-1}-c)^\gamma \theta(R_{k-1}-c) \right]^{1/\gamma}
\end{equation}
Using $x_k=-\ln(1-u_k)$ ($x_k$ becomes an exponentially distributed number with mean $1$):
\begin{equation}
\label{eqn:weibulltail_sample}
    R_k=\left[x_k + (R_{k-1}-c)^\gamma \theta(R_{k-1}-c) \right]^{1/\gamma}
\end{equation}
Equation (\ref{eqn:weibulltail_sample}) is nothing but the rule for sampling the tail of a Weibull distribution. Similar rules can be written for treatable bounded distributions $F(x)=1-(1-x)^\nu$. For all other distributions one has to rely on specialized methods. Given the record values $R_1, R_2, \dots$ we can sample the ages $n_1, n_2, \dots$ by using the Geometrical distribution. Indeed if we marginalize $R$ from equation (\ref{eqn:Rn_cond_prop}) we get the conditional age distribution:
\begin{equation}
    \pi(n|R') = \begin{cases}
 	(1-F(R'-c)) F(R'-c)^{n-1} & R'>c \\
 	\delta_{n,1} & R'<c
    
    \end{cases}
\end{equation}
Hence, given a record $R_{k-1}$, the age between $R_{k-1}$ and the next record $R_k$ is geometrically distributed.
\end{document}